\documentclass[utf8]{frontiersSCNS}
\usepackage{url,hyperref,lineno,microtype,subcaption}
\usepackage[onehalfspacing]{setspace}
\usepackage{xcolor}
\usepackage{lscape}

\def\keyFont{\fontsize{8}{11}\helveticabold }
\def\firstAuthorLast{Michel \& Mugrauer}
\def\Authors{K.-U. Michel\,$^{1,*}$ \& M. Mugrauer\,$^{1}$ }
\def\Address{$^{1}$Astrophysikalisches Institut und Universit\"{a}ts-Sternwarte Jena, Schillerg\"{a}{\ss}chen 2, D-07745 Jena, Germany}
\def\corrAuthor{Corresponding Author: K.-U. Michel}
\def\corrEmail{kai-uwe.michel@uni-jena.de}
\begin{document}
\onecolumn
\firstpage{1}

\title[Stellar companions of exoplanet hosts]{Search for (sub)stellar companions of exoplanet hosts by exploring the second ESA-Gaia data release}

\author[\firstAuthorLast ]{\Authors}
\address{}
\correspondance{}
\extraAuth{}
\maketitle

\begin{abstract}
We present the latest results of an ongoing multiplicity survey of exoplanet hosts, which was initiated at the Astrophysical Institute and University Observatory Jena, using data from the second data release of the ESA-Gaia mission. In this study the multiplicity of 289 targets was investigated, all located within a distance of about 500\,pc from the Sun. In total, 41 binary, and 5 hierarchical triple star systems with exoplanets were detected in the course of this project, yielding a multiplicity rate of the exoplanet hosts of about 16\,\%. A total of 61 companions (47 stars, a white dwarf, and 13 brown dwarfs) were detected around the targets, whose equidistance and common proper motion with the exoplanet hosts were proven with their precise Gaia DR2 astrometry, which also agrees with the gravitational stability of most of these systems. The detected companions exhibit masses from about 0.016 up to 1.66\,M$_\odot$ and projected separations in the range between about 52 and 9555\,au.

\tiny\keyFont{\section{Keywords:} Multiple Stars, White Dwarfs, Brown Dwarfs, Exoplanets, ESA-Gaia DR2}
\end{abstract}

\section{Introduction}

Since the detection of the first planet orbiting a star other than the Sun, several thousands of these exoplanets have been discovered by various detection techniques. While the majority of stars are members of multiple star systems \citep{Duchene2013}, most of the exoplanet host stars are single stars. Nevertheless several multiple star systems hosting exoplanets, could already be revealed by previous multiplicity studies using seeing limited or high contrast AO imaging observations \citep[see e.g.][]{mugrauer2014, mugrauer2015}. In order to explore the effects of the presence of stellar companions on the formation process and orbital evolution of exoplanets, a survey was initiated at the Astrophysical Institute and University Observatory Jena  \citep[described in detail by][]{mugrauer2019} to identify and characterize companions of exoplanet host stars, detected in the second data release of the European Space Agency (ESA) Gaia mission \citep[Gaia DR2 from hereon,][]{gaia2018}. Furthermore, in \cite{mugrauer2020} a comparable investigation was carried out among potential exoplanet host stars, identified by the TESS mission \citep{ricker2015}. The study, whose results are presented here, is the third work in the context with \cite{mugrauer2019}. The following section gives a detailed description of this study, and the detected companions and their derived properties are presented in the third section of this paper.

\section{Gaia DR2 search for (sub)stellar companions of exoplanet hosts}

The Gaia DR2, is based on data taken by the Gaia spacecraft in the first 22 months of its mission and contains $1.7$ billion detected sources up to a limiting magnitude of $G=21$\,mag. For 1.3 billion sources a five parameter astrometric solution could be derived, i.e. beside their equatorial coordinates ($\alpha$, $\delta$), also the parallax $\pi$ and proper motion ($\mu_{\alpha}cos(\delta)$, $\mu_{\delta}$) of these sources were determined. Furthermore, for about 88 million detected objects estimates of their G-band extinction and effective temperature are listed in the Gaia DR2, determined by the Priam algorithm, which is part of the astrophysical parameters inference system \citep[Apsis, see][]{bailer-jones2013} in the Gaia data processing.

Using Gaia DR2 data \cite{mugrauer2019} already explored the multiplicity of all exoplanet host stars, whose exoplanets were detected either by photometric transit observations, radial-velocity (RV), or astrometric measurements, and were listed in the \texttt{Extrasolar Planets Encyclopaedia}\footnote{Online available at: \url{http://exoplanet.eu/}} \citep[\texttt{EPE} from hereon,][]{schneider2011} by mid of October 2018. The study, presented in this paper, complements this survey by investigating the multiplicity of the exoplanet hosts (stars but also brown dwarfs), whose planets were indirectly detected either via RV measurements or transit observations in the range of time between mid of October 2018 until end of September 2020, as well as all exoplanet hosts, known so far, with planets, which were directly detected by imaging observations. At the end of September 2020 the \texttt{EPE} lists about 4350 exoplanets, and about 400 of them were detected around the hosts studied in this work.

\begin{figure}[h!]
\begin{center}
\resizebox{\hsize}{!}{\includegraphics{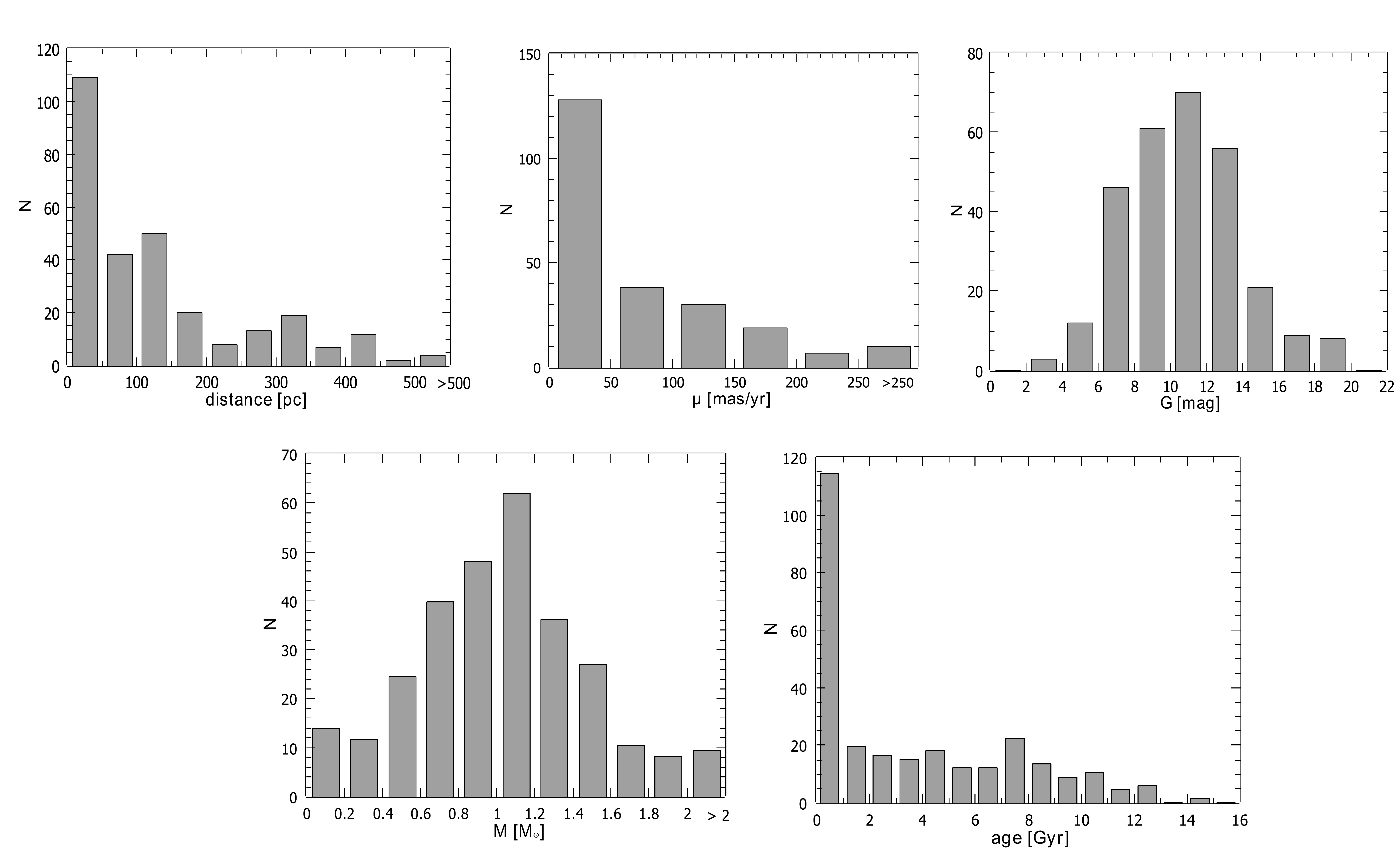}}
\caption{The histograms of the individual properties of all targets of this study.}
\label{Fig_TargetProps}
\end{center}
\end{figure}

(Sub)stellar Companions are expected to be located at the same distance to the Sun as the exoplanet hosts and form common proper motion pairs with them, in particular wide companions with projected separations of hundreds and thousands of au, i.e. the typical targets of this study. Hence, in order to clearly detect such companions and to prove the equidistance of these objects and the exoplanet hosts, in this study we have taken into account only Gaia DR2 sources with an accurate five parameter astrometric solution, i.e. which exhibit precise measurements of their parallax ($\pi/\sigma(\pi) > 3$) and proper motion ($\mu/\sigma(\mu) > 3$). Thereby, sources with negative parallaxes are neglected. As in the Gaia DR2 a parallax uncertainty of 0.7\,mas is reached for faint sources down to $G = 20$\,mag, the survey is furthermore constrained to exoplanet hosts, which are located within a distance of 500\,pc around the Sun (i.e. $\pi > 2$\,mas), to assure $\pi/\sigma(\pi) > 3$ even for the faintest companions, detectable in this survey. This distance constraint is slightly relaxed to $\pi + 3\sigma(\pi)\gtrapprox2$\,mas, i.e. taking into account also the parallax uncertainty of the hosts. By the end of September 2020, in total 289 exoplanet hosts are listed in the \texttt{EPE}, which fulfill this distance constraint, and hence are selected as targets for this study. The properties of all targets are summarized in Tab.\,\ref{Tab_TargetProps} and their histograms are illustrated in Fig.\,\ref{Fig_TargetProps}. On average, the targets are solar like stars most frequently found within 150\,pc around the Sun, which exhibit proper motions in the range between about 2 and 10400\,mas/yr, and G-band magnitudes from about 3.7 to 20.8\,mag. In particular, the sub-sample of direct imaging exoplanet hosts emerges as a peak in the age distribution at young ages, as all these targets are typically younger than 0.1\,Gyr, in contrast to hosts of RV and transiting exoplanets, which are older than 1\,Gyr in general.

The companion search radius, applied in this project around the selected targets, is limited to a maximal projected separation of 10000\,au, which guarantees that the majority of wide companions of the exoplanet hosts are detectable in this study, as described by \cite{mugrauer2019}. This upper separation limit results in an angular search radius around the targets of $r [\text{arcsec}] = 10 \pi[\text{mas}]$. Within this radius around the targets the companionship of all sources, listed in the Gaia DR2 with an accurate five parameter astrometric solution was investigated. For the verification of the equidistance of all detected sources with the associated exoplanet hosts, the difference $\Delta \pi$ between their parallaxes was calculated, taking into account also the excess noise of their astrometric solutions. Common proper motion of the detected sources and the targets was checked with the precise Gaia DR2 proper motions of the exoplanet hosts $\mu_{PH}$ and the sources $\mu_{Comp}$. In addition, we have also derived for all sources the differential proper motion: $\mu_{rel} = |\mu_{PH} - \mu_{Comp}|$, which yields the common proper motion index ($\textrm{cpm-index} = |\mu_{PH}+\mu_{Comp}| / \mu_{rel}$), which characterizes the degree of common proper motion of the detected sources and the exoplanet hosts.

Following the companion identification procedure ($sig$-$\Delta\pi \leq 3$ \& $\textrm{cpm-index}\geq 3$), as defined by \cite{mugrauer2019} the majority of all sources ($>$99.88\,\%), detected within the applied search radius around the targets, can clearly be excluded as companions, as they are either not located at the same distances as the exoplanet hosts and/or do not share a common proper motion with them. In contrast, for 61 detected objects their companionship with the targets could clearly be proven with their precise Gaia DR2 astrometry. For all these companions we have determined their relative astrometry to the exoplanet hosts (angular separation $\rho$, and position angle $PA$), as well as their projected separation $sep$, derived with their angular separation and the parallax of the targets.

The absolute G-band magnitude of all companions was derived from their apparent G-band photometry, the parallax of the associated exoplanet hosts, as well as their Apsis-Priam G-band extinction estimate, all listed in the Gaia DR2. If there was no extinction estimate given for a companion, the extinction estimate of the exoplanet host was used instead or if not available, its extinction estimate, listed in the \texttt{StarHorse} catalog \citep{anders2019}. In the case that no G-band extinction is available at all it was derived from V-band extinction measurements of the exoplanet hosts, listed either in the \texttt{VizieR} data base\footnote{Online available at: \url{https://vizier.u-strasbg.fr/}} \citep{vizier2000} or in the literature, adopting $A_G / A_V = 0.77$, as described by \citet{mugrauer2019}.

\begin{figure}[h!]
\begin{center}
\resizebox{\hsize}{!}{\includegraphics{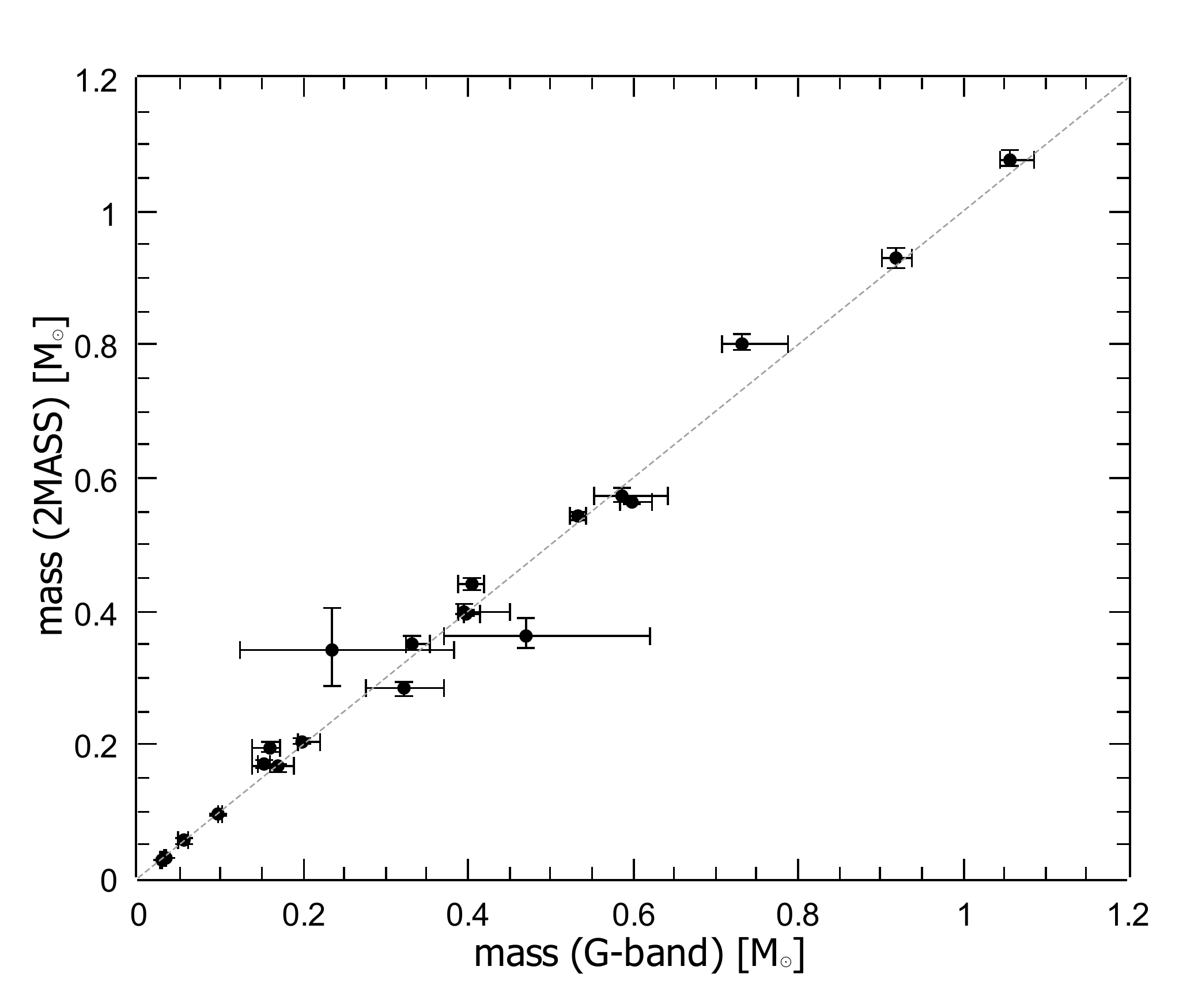}}
\caption{Comparison of the mass of the detected companions, derived from their G-band and infrared 2MASS photometry.}
\label{Fig_2mass}
\end{center}
\end{figure}

The masses and effective temperatures of all detected companions were determined from their derived absolute G-band magnitudes using the evolutionary models of (sub)stellar objects from \cite{baraffe2015}, as well as the ages of the exoplanet hosts, as listed in the \texttt{EPE}. Thereby, we adopt the same age for the planet hosts and their companions. We determined the masses and effective temperatures of the companions via interpolation of the model grid with the age closest to that of the exoplanet hosts.
For verification of the obtained results the properties of the companions derived from their G-band magnitudes were compared with those, determined from the near-infrared photometry, taken from the 2MASS Point Source catalogue \citep{skrutskie2006}, if available. For the near-infrared extinction we have used the relations: $A_{Ks}/A_V = 0.12$, $A_H / A_V = 0.17$, and $A_J/A_V=0.26$, as described in \cite{mugrauer2019}. A graphical comparison of the masses obtained from the G-band and the 2MASS photometry are shown in Fig.\,\ref{Fig_2mass}. The identity is illustrated as grey dashed line in this figure. For all companions the derived masses agree well with each other, with deviations that remain below the $3\sigma$ level (the same holds also for the temperature estimates not shown here). Objects, whose masses were determined by extrapolation from the used model grids as such as those with bad quality (quality flags all but A) or contaminated 2MASS photometry were excluded in this comparison.

Eventually for all companions, which were detected in this study, we have estimated their escape velocity $\mu_{esc}\,[mas\,yr^{-1}] = 2\pi\sqrt{2M \pi^{3}_{PH} / \rho}$ with their angular separation $\rho$ and the parallax of the associated exoplanet hosts both in the unit of milli-arcsec (mas), as well as the total mass $M$ of the system (in the unit $M_\odot$), i.e. the sum of the mass of the companions, derived as describe above, and the mass of the associated exoplanet hosts, taken from the \texttt{EPE}. This estimation can be considered as an upper limit of the escape velocity as the projected separation is smaller than the physical separation of the objects.

\section{Detected companions of exoplanet hosts}

The Gaia astro- and photometry of all exoplanet hosts and their companions, detected in this study, are listed in Tab.\,\ref{Tab_Gaia_AstroPhoto}. The derived properties of the companions are summarized in Tab.\,\ref{Tab_CompAstro} to \ref{Tab_gravtest}. In all tables the exoplanet host systems or the companions are sorted by their right ascension. The used identifier of the targets corresponds either to the one used in the \texttt{EPE} or is a slightly abbreviated version of it. In contrast to the planet definition used by the \texttt{EPE}, in which substellar objects below 60\,$M_{Jup}$ are defined as exoplanets, we follow here the planet definition based on the deuterium burning limit \citep[as described e.g. by][]{Basri2000}, i.e. all substellar objects below 13\,$M_{Jup}$ are classified as exoplanets, while more massive objects below the substellar/stellar mass limit (at about 0.072\,$M_{Jup}$ for solar metallicity) as brown dwarfs, respectively. Thereby the given masses of the exoplanets, detected by radial velocity measurements, correspond to minimum-masses ($M \sin(i)$) due to the unknown orbital inclination, while masses of direct imaging planets are usually derived from their spectrophotometry with evolutionary models.

In Tab.\,\ref{Tab_Gaia_AstroPhoto} for each exoplanet host and its detected co-moving companion(s) their Gaia DR2 parallax $\pi$, proper motion in right ascension and declination ($\mu_\alpha\cos(\delta)$ \& $\mu_\delta$), astrometric excess noise (epsi) with its significance (sig-epsi), apparent G-band magnitude, as well as the used Apsis-Priam G-band extinction estimate $A_G$ are listed. In the case that the G-band extinction was taken from the \texttt{StarHorse} catalog this is indicated with the \texttt{SHC} flag, or with the $\maltese$ flag if the G-band extinction was derived from V-band extinction measurements, either listed in the \texttt{VizieR} database or from the literature. In this table the exoplanet hosts are indicated with \texttt{*}, and known spectroscopic binary stars among them with \texttt{(SB)}.

Table\,\ref{Tab_CompAstro} lists for each detected companion its angular separation ($\rho$) and position angle ($PA$) to the associated exoplanet host, which were determined with the Gaia DR2 astrometry of the objects for the (Gaia reference) epoch 2015.5. The relative astrometry of the companions exhibits an uncertainty on average of 0.3\,mas in angular separation, and 0.002\,$^{\circ}$ in position angle, respectively. In the following columns of Tab.\,\ref{Tab_CompAstro} we list the parallax difference ($\Delta \pi$) with its significance (in brackets calculated by taking into account also the Gaia astrometric excess noise\footnote{The astrometric excess noise is conservatively considered here as additional parallax uncertainty of the source.}) between the exoplanet hosts and their detected companions, their differential proper potion $\mu_{rel}$ with its significance, and the cpm-index of all systems. The precise Gaia DR2 astrometry proves the equidistance (sig-$\Delta \pi < 2.3\,\sigma$, average value of $0.5\,\sigma$) and common proper motion ($\textrm{cpm-index} > 6$, $\textrm{average cpm-index}=118$) of the exoplanet hosts and their companions. If these companions are not listed yet as companion(-candidates) in the Washington Double Star Catalog \citep[WDS from hereon,][]{mason2001} this is indicated with the $\bigstar$ flag in last column of Tab.\,\ref{Tab_CompAstro}. In the case that the companion is not listed in the WDS but was reported in literature before, additional information is given in the notes of this table.

In Tab.\,\ref{Tab_CompProps} beside the equatorial coordinates ($\alpha$, $\delta$ both for epoch 2015.5) of all detected companions, their derived absolute G-band magnitude $M_G$, projected separation $sep$ to the associated exoplanet host (relative uncertainty about 1\,\%, on average), mass, and effective temperature $T_{eff}$ are summarized. The flags listed in the last column of this table are defined as follows:

\begin{itemize}
\item \texttt{PRI}: An Apsis-Priam temperature estimate is available for the detected companion, which could be compared with the effective temperature of the companion, derived from its absolute G-band photometry using the \cite{baraffe2015} models.
\item \texttt{2MA}: The companion is listed in the 2MASS Point Source catalogue.
\item \texttt{BPRP}: The $G_{BP}-G_{RP}$ color of the exoplanet host and of the detected companion is listed in the Gaia DR2, hence a color comparison was feasible.
\item \texttt{EXT}: Because of its brightness the companion exceeds the magnitude range of the \cite{baraffe2015} evolutionary models. Therefore, the properties of the companion were estimated via extrapolation from the two brightest sources of the used model isochrone.
\item \texttt{WD}: The detected companion is a white dwarf.
\item \texttt{BD}: The detected companion is a brown dwarf.
\end{itemize}

Finally, in Tab.\,\ref{Tab_gravtest} we summarize all those detected companions, whose differential proper motion $\mu_{rel}$ significantly exceeds their expected escape velocity $\mu_{rel}$. Companions, which are already known to be members of hierarchical triple star systems, are indicated with the flag \texttt{***} in the last column of this table.

\begin{figure}[h!]
\begin{center}
\resizebox{\hsize}{!}{\includegraphics{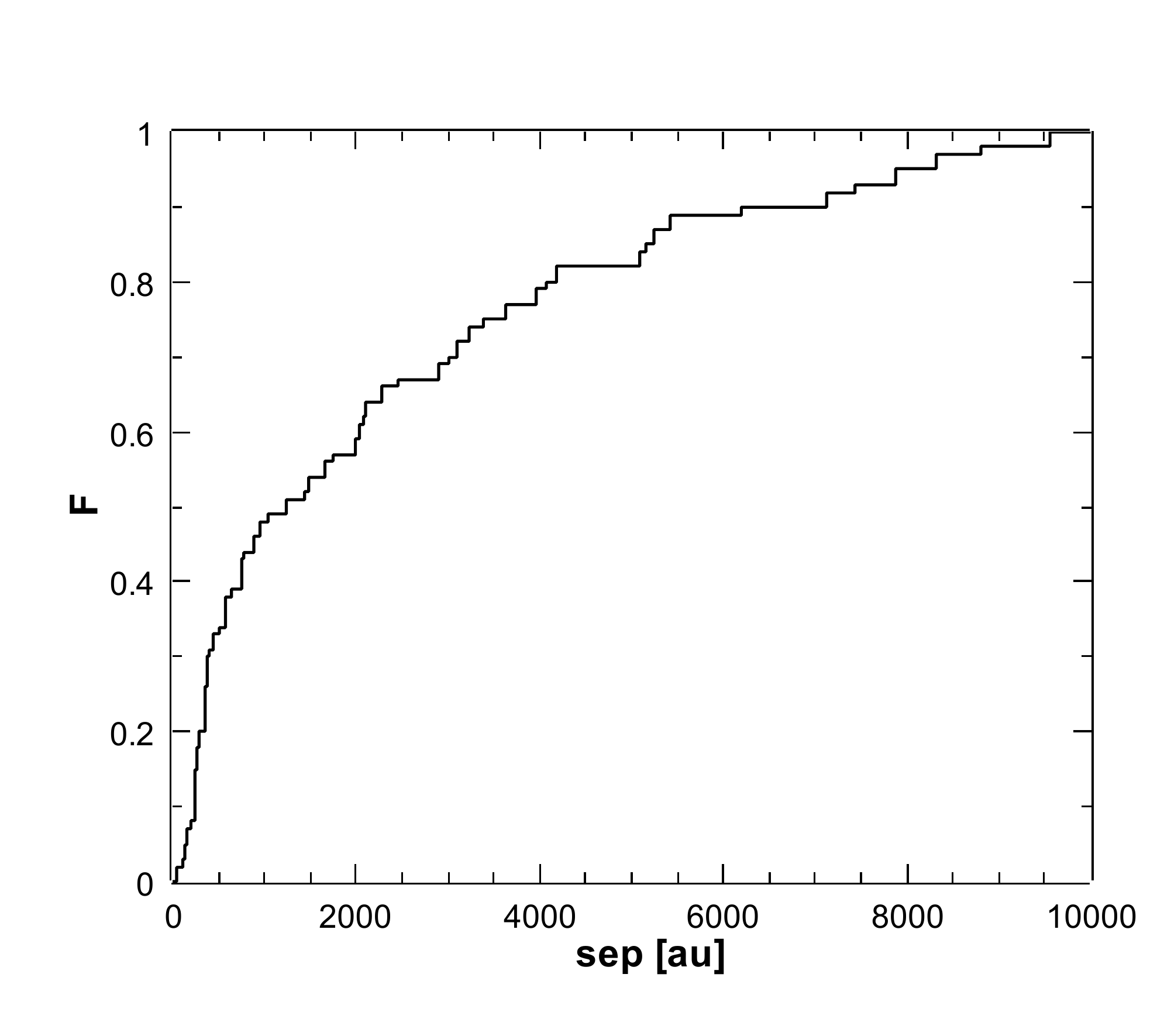}}
\caption{The cumulative distribution function of the projected separation ($sep$) of all detected companions to the associated exoplanet hosts.}
\label{Fig_compsep}
\end{center}
\end{figure}

\begin{figure}[h!]
\begin{center}
\resizebox{\hsize}{!}{\includegraphics{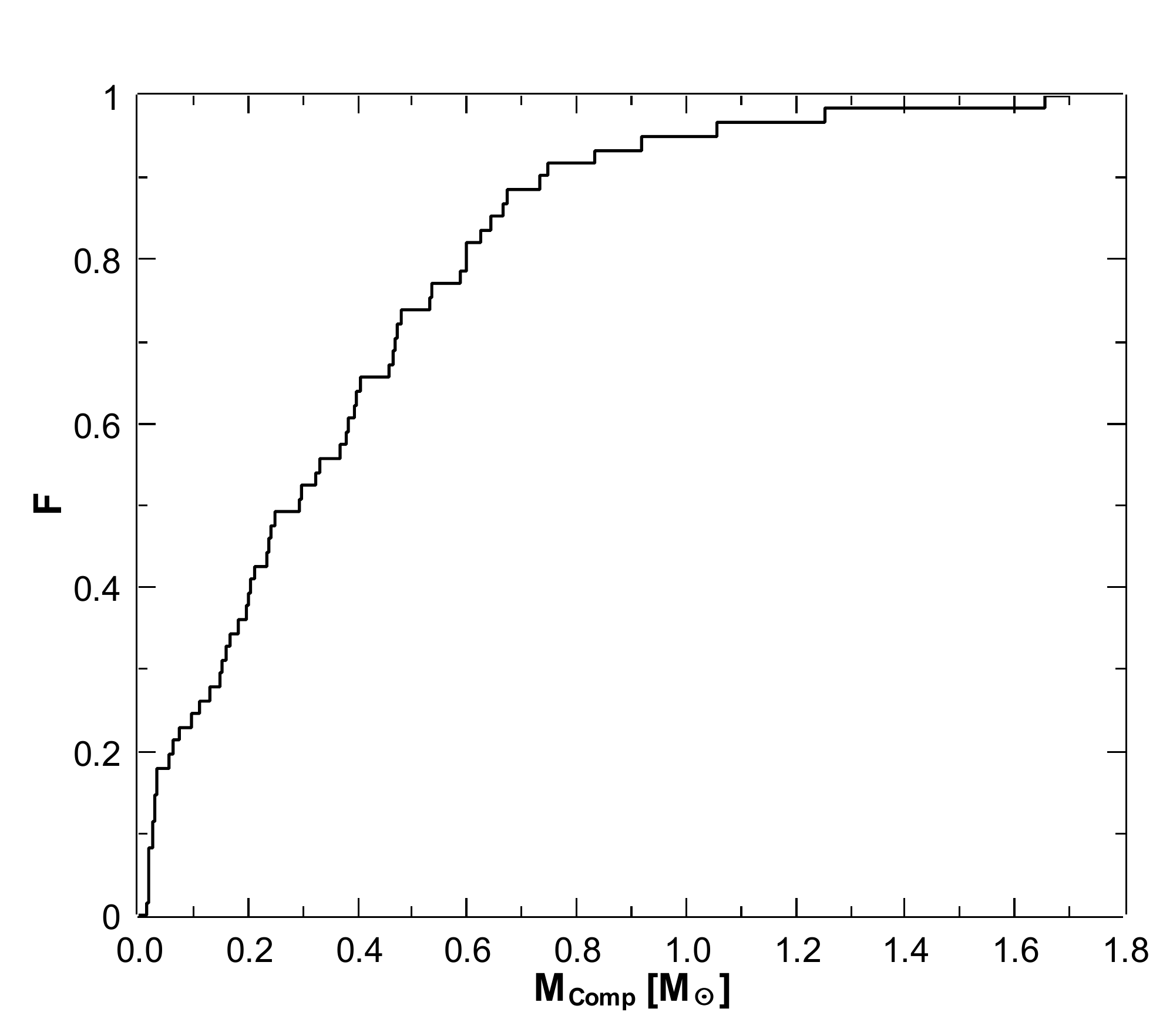}}
\caption{The cumulative distribution function of the mass of all companions, detected in this study.}
\label{Fig_compmass}
\end{center}
\end{figure}

\begin{figure}[h!]
\begin{center}
\resizebox{\hsize}{!}{\includegraphics{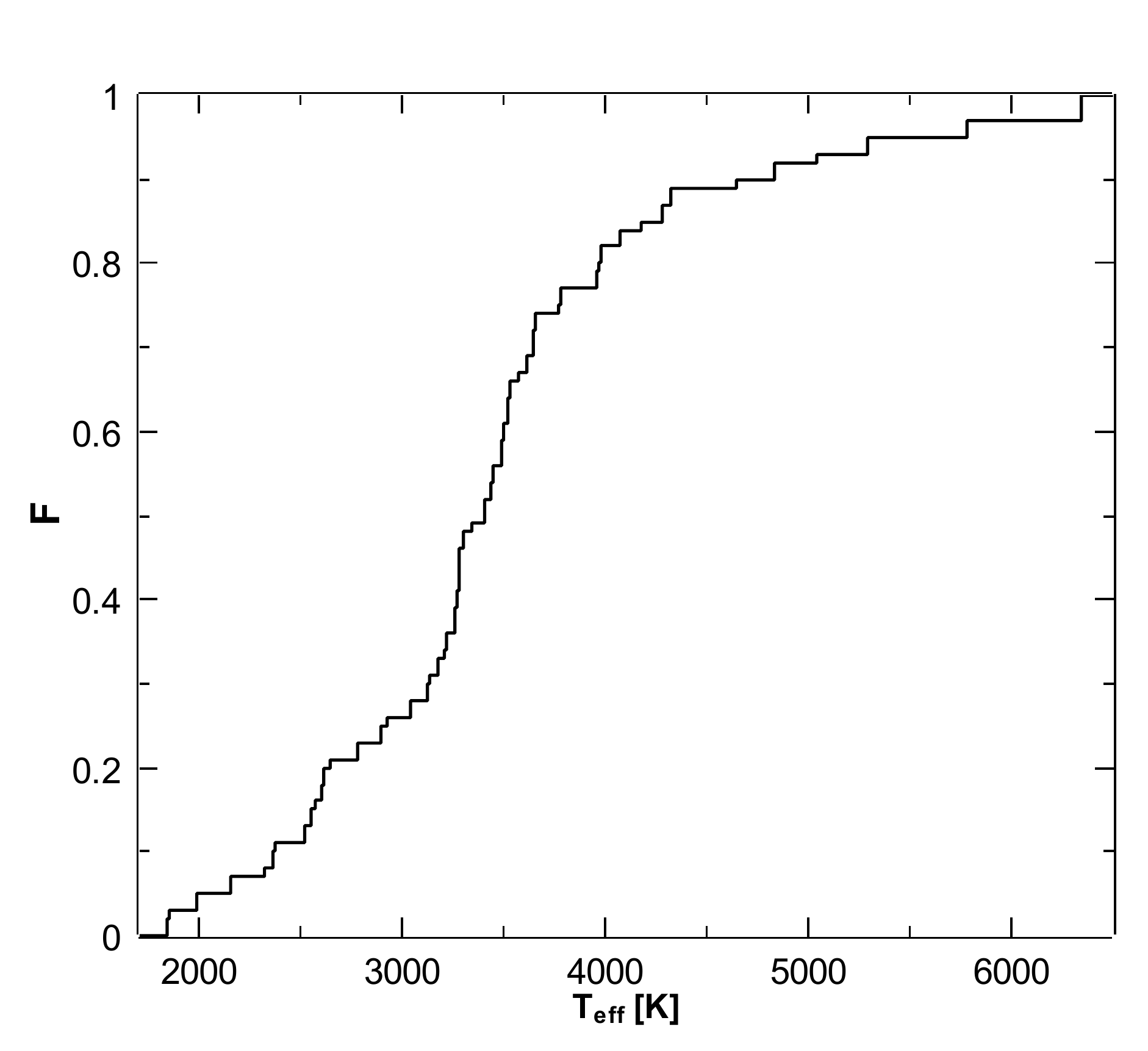}}
\caption{The cumulative distribution function of the effective temperature of all detected companions.}
\label{Fig_compteff}
\end{center}
\end{figure}

\begin{figure}[h!]
\begin{center}
\resizebox{\hsize}{!}{\includegraphics{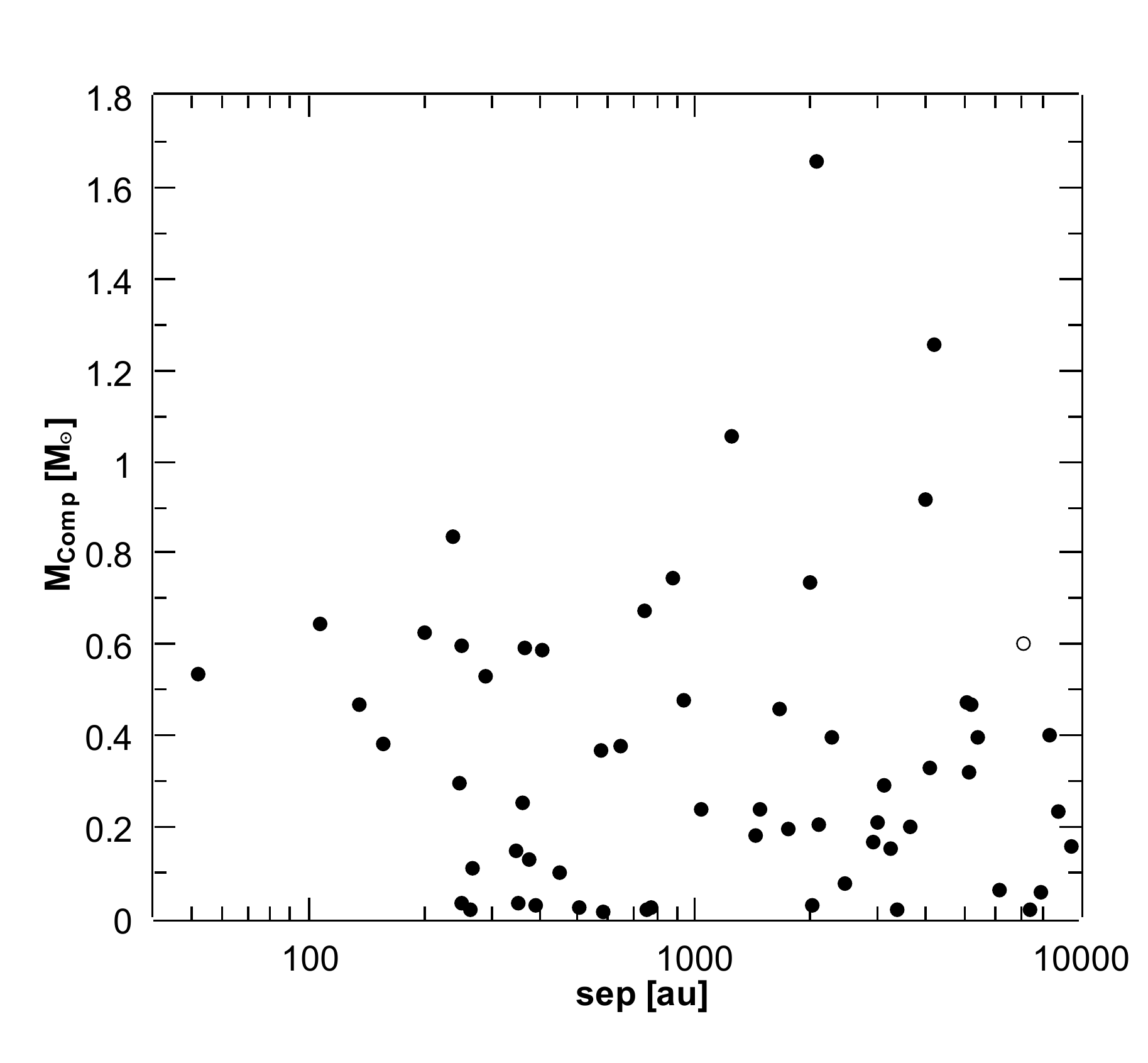}}
\caption{The mass of all companions, detected in this study, plotted over their projected separation ($sep$) to the associated exoplanet hosts. The white dwarf companion HIP\,38594\,B, is illustrated as open circle.}
\label{Fig_sepmass}
\end{center}
\end{figure}

Among all 289 targets, whose multiplicity was investigated in the study, whose results are presented in this paper, 41 binary and 5 hierarchical triple star systems with exoplanets were identified. This yields a multiplicity rate of the targets of $16\pm$2\%, very well consistent with the multiplicity rate of exoplanet host stars of $15\pm1$\,\%, reported before by \cite{mugrauer2019}.
This is as expected, as the sensitivities of the two surveys should agree well with each other, as the brightness and mass of their targets match, and the distance of the targets from this survey is on average about 40\,\% smaller than that of the targets from \cite{mugrauer2019}, resulting in a reduction in the distance modulus of only about 1\,mag. In total, 61 companions (48 stars and 13 brown dwarfs) could be detected in the Gaia DR2 around the targets. The detected substellar companions are all listed as exoplanets in the \texttt{EPE}. The cumulative distribution functions of the derived properties (projected separation, mass and effective temperature) of theses companions, are illustrated in Fig.\,\ref{Fig_compsep}, \ref{Fig_compmass}, and \ref{Fig_compteff}. The separation-mass diagram of the companions is shown in Fig.\,\ref{Fig_sepmass}. As described above, the accurate Gaia DR2 astrometry proves the equidistance and common proper motion of all detected companions with the associated exoplanet hosts, and for the majority of these companions their differential proper motion to the exoplanet hosts is slower than their estimated escape velocity, facts that are expected for gravitationally bound systems. In contrast, the differential proper motion of the companions, which are listed in Tab.\,\ref{Tab_gravtest}, exceeds their estimated escape velocity, possibly indicating a higher degree of multiplicity.\footnote{Additional close companions either of the exoplanet hosts or of the companions force these objects on close orbits with high orbital velocities around a common barycenter that could induce the observed high differential velocities.}
Indeed, one of these companions (51\,Eri\,BC) is already known to be a close binary itself. The remaining 2 companions and their primaries are promising targets for follow-up observations to check their multiplicity status e.g. with high contrast AO imaging observations.

All detected companions exhibit projected separations to the associated exoplanet hosts in the range between 52 and 9555\,au (average separation of about 2310\,au). The highest companion frequency is found at projected separations between about 240 and 400\,au and half of all companions are located at projected separations below about 1240\,au. The closest detected companion is K2-288\,A, which is separated from the exoplanet host stars K2-288\,B by 52\,au, and it is the only companion identified in this study within a projected separation of 100\,au. The masses of the companions range between 0.016 and 1.66\,$M_\odot$ (average mass of 0.36\,$M_\odot$) and companions are found most frequently in the substellar mass regime between 0.016 up to 0.033\,$M_\odot$, while more massive companions are detected at a lower but constant frequency up to about 0.7\,$M_\odot$, and only about 10\,\% of all the detected companions exhibit masses larger than 0.7\,$M_\odot$. The companions exhibit effective temperatures in the range between about 1850 and 6350\,K (average temperature of about 3400\,K), which corresponds to spectral types of L3 to F6 (M3, on average), according to the $T_{eff} - SpT$ relation\footnote{Online available at: \url{http://www.pas.rochester.edu/~emamajek/EEM_dwarf_UBVIJHK_colors_Teff.txt}} from \citep{pecaut2013}.

In general the effective temperature of the detected companions, determined with their derived absolute G-band magnitude, using the evolutionary \cite{baraffe2015} models, agree well with their Gaia DR2 Apsis-Priam temperature estimate (if available) with a characteristic deviation of about $\pm$350\,K, consistent with the typical uncertainty of the different temperature estimates, which is in the order of about 330\,K. Only in the case of HIP\,38594\,B the temperature estimate, based on the absolute G-band photometry of the companion significantly deviates by more than 2300\,K from its Apsis-Priam temperature estimate, which is also about 900\,K higher than the one of the associated exoplanet host star HIP\,38594\,A. Furthermore, the companion appears bluer ($\Delta(G_{BP}-G_{RP}) = -0.669\pm0.004$\,mag) than its primary although it is about 7\,mag fainter in the G-band than the exoplanet host star. The intrinsic faintness and high temperature of HIP\,38594\,B clearly indicates that this companion is a white dwarf. This conclusion is consistent with the results of \cite{subasavage2008}, who have already classified the companion spectroscopically as a white dwarf, and have denote it as WD\,0751-252. For this degenerated companion we adopt here a mass of about 0.6\,$M_\odot$.

\begin{figure}[h!]
\begin{center}
\resizebox{\hsize}{!}{\includegraphics{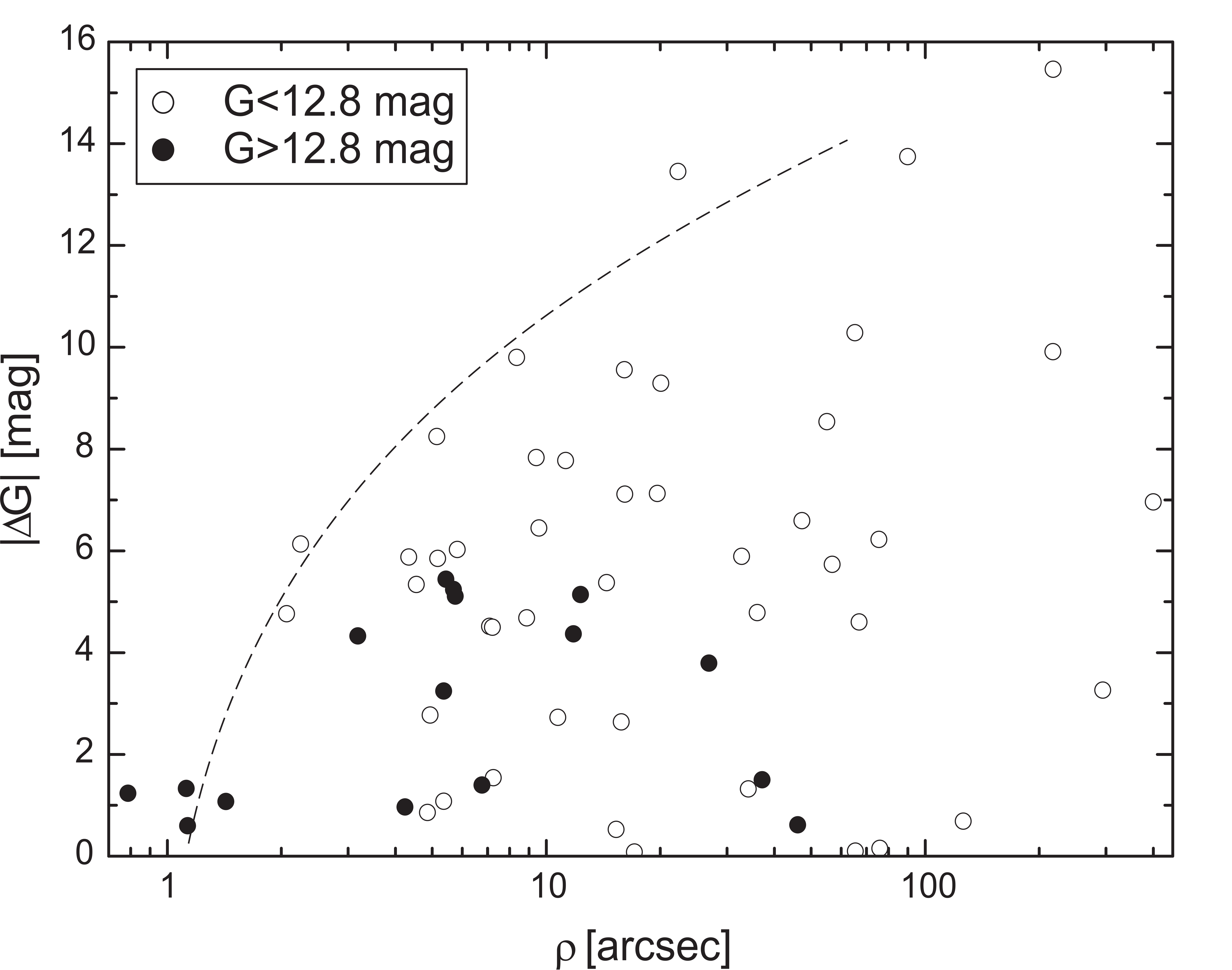}}
\caption{The G-band magnitude difference of all detected companions, plotted over their angular separation to the associated exoplanet hosts.}
\label{Fig_detectionlimit}
\end{center}
\end{figure}

In Fig.\,\ref{Fig_detectionlimit} the G-band magnitude difference of all detected companions to the associated exoplanet hosts is plotted versus their angular separation. For comparison we show as dashed line in this figure the estimate of the Gaia detection limit, reported by \cite{mugrauer2019} which was further constrained by \cite{mugrauer2020}. Companions of exoplanet hosts brighter than 12.8\,mag are plotted as open circles those of hosts, which are fainter than that magnitude limit, as filled black circles, respectively. A magnitude difference of about 5\,mag is reached at an angular separation of about 2\,arcsec, consistent with the estimate of the Gaia detection limit, determined by \cite{mugrauer2019}. Only two companions significantly exceed the limit estimate, namely K2-288\,A at an angular separation of about 0.8\,arcsec with $\Delta G \sim 1.2$\,mag and HIP\,77900\,B, at 22.3\,arcsec with $\Delta G\sim 13.5$\,mag. While K2-288\,A is a companion of a target fainter than $G=12.8$\,mag for which Gaia reaches a higher sensitivity at angular separations slightly below 1\,arcsec \cite[up to 3\,mag, as described by][]{mugrauer2020} the detection of HIP\,77900\,B indicates that the given limit estimate might be too conservative at angular separations beyond about 20\,arcsec.

\section{Summary and Outlook}

The study, presented here, is a continuation of a survey, which was initiated at the Astrophysical Institute and University Observatory Jena, to investigate the multiplicity status of exoplanet hosts and to characterize the properties of their detected (sub)stellar companions, using accurate Gaia astro- and photometry. In this paper the multiplicity of 289 exoplanet hosts was explored and (sub)stellar companions were detected around 60 targets. The companionship of these objects with the exoplanet hosts could be proven with their accurate Gaia DR2 astrometry (equidistance, common proper motion, and differential proper motion smaller than the expected escape velocity). The mass and effective temperature of all companions were determined with their derived absolute G-band photometry and the \cite{baraffe2015} evolutionary models of (sub)stellar objects. In total, 61 companions (beside 48 stellar companions, among them the white dwarf HIP\,38594\,B, also 13 brown dwarfs) were detected in this project, and 14 of these objects are neither listed in the WDS as companion(-candidate)s of the targets nor were described in the literature before. A total of 41 binary and 5 triple star systems with exoplanets, were identified in this study, yielding a multiplicity rate of the targets of about 16\,\%, which is very well consistent with the multiplicity rate of exoplanet host stars, reported by \cite{mugrauer2019}. Following the standard procedure of our survey, all detected companions and their derived properties will be made available online in the \texttt{VizieR} database. The survey, whose latest results are presented here, is an ongoing project as more and more exoplanet hosts are detected by different planet detection methods, whose multiplicity status needs to be investigated. Furthermore, there are sources, listed in the Gaia DR2, within the applied search radius around the targets, which still lack a five parameter astrometric solution. Hence, further companions of the exoplanet hosts, investigated here, should exist, whose companionship can be proven with accurate astrometric measurements, provided by future data releases of the ESA-Gaia mission, e.g. the Gaia EDR3, planed to be published end of 2020.

The results of this survey, which is mainly sensitive for wide companions of exoplanet hosts, combined with those of our currently ongoing large high contrast imaging surveys (sensitive for close companions), carried out with SPHERE/VLT and AstraLux/CAHA \citep[first results are already published  e.g. by][]{ginski2020} will yield a complete characterization of the multiplicity status of the observed targets. This will eventually allow to draw conclutions on the impact of the stellar multiplicity on the formation process of planets and the evolution of their orbits.

\section*{Acknowledgments}

We thank the two anonymous referees for their helpful and constructive comments on the manuscript.

We made use of data from:

(1) the \texttt{Simbad} and \texttt{VizieR} databases, both operated at CDS in Strasbourg, France.

(2) the European Space Agency (ESA) mission Gaia (\url{https://www.cosmos.esa.int/gaia}), processed by the Gaia Data Processing and Analysis Consortium (DPAC, \url{https://www.cosmos.esa.int/web/gaia/dpac/consortium}). Funding for the DPAC has been provided by national institutions, in particular the institutions participating in the Gaia Multilateral Agreement.

(3) the Two Micron All Sky Survey, which is a joint project of the University of Massachusetts and the Infrared Processing and Analysis Center/California Institute of Technology, funded by the National Aeronautics and Space Administration and the National Science Foundation.

\bibliographystyle{frontiersinSCNS_ENG_HUMS}
\bibliography{michel}

\newpage

\section{Tables}

\begin{table}[h!] \caption{The properties of all targets of this study. The corresponding histograms are shown in Fig.\,\ref{Fig_TargetProps}.}
\begin{center}
\begin{tabular}{cccccc}
\hline\hline
    &  Distance          & $\mu$           & G          & age        & mass \\
    &  $[$pc$]$          & $[$mas/yr$]$    & $[$mag$]$  & $[$Gyr$]$  & $[M_\odot]$\\
\hline\hline
min & 1.8                & 1.7             & 3.7        & 0.001      & 0.016      \\
\hline
max & 586                & 10394           & 20.8       & 14.9       & 20         \\
\hline
ave & 137                & 270             & 10.8       & 3.5        & 1.1        \\
\hline
med & 94                 & 65              & 10.7       & 2.1        & 1.0        \\
\hline\hline
\end{tabular}
\end{center}
\label{Tab_TargetProps}
\end{table}

\begin{landscape}
\begin{table*} \caption{Gaia astro- and photometry of all exoplanet hosts and their companions, detected in this study.}
\begin{small}
\begin{center}
\begin{tabular}{lcccccccc}
\hline\hline
Name                & $\pi$         & $\mu_{\alpha}cos(\delta)$ & $\mu_{\delta}$ & $epsi$ & $sig$- & G  & $A_{G}$ \\
                    & [mas]              & [mas/yr]                     & [mas/yr]       & [mas]  & $epsi$ & [mag] & [mag]   \\
\hline\hline
HD\,1160\,A*        & \,\,\,$7.9417\pm0.0764$   & $\,\,\,\,20.089\pm0.138$   &$-14.575\pm0.099$  & $0.121$ & $6.0$ & $\,\,\,7.1074\pm0.0003$  & $0.1347_{-0.0968}^{+0.1300}$\\
HD\,1160\,C         & \,\,\,$6.9946\pm0.2739$   & $\,\,\,\,20.605\pm0.333$   &$-16.215\pm0.311$  & $0.739$ & $37$ & $15.3505\pm0.0207$ & \\\hline
Gliese\,49\,A*      & $101.4650\pm0.0335\,\,\,$ & $\,\,731.135\pm0.041$  &$\,\,\,\,90.690\pm0.048$   & $-$     & $-$    & $\,\,\,8.6628\pm0.0007$  & $0.6030_{-0.4095}^{+0.2220}$\\
Gliese\,49\,B       & $101.6371\pm0.0806\,\,\,$ & $\,\,730.740\pm0.163$  &$\,\,\,\,86.352\pm0.225$   & $0.190$ & $13$ & $11.9238\pm0.0033$ & \\\hline
HD\,8326\,A*        & $32.5591\pm0.0466$  & $-58.470\pm0.120$  &$-224.887\pm0.064\,\,\,\,$ & $-$     & $-$    & $\,\,\,8.4749\pm0.0004$  & \\
HD\,8326\,B         & $32.4362\pm0.0589$  & $-57.577\pm0.156$  &$-224.122\pm0.088\,\,\,\,$ & $0.347$ & $27$ & $14.2066\pm0.0006$ & $0.2940_{-0.0438}^{+0.2446}$\\\hline
HD\,13167\,A*       & \,\,\,$6.6859\pm0.0485$   & $\,\,\,\,43.770\pm0.077$   &$-38.126\pm0.079$  & $-$     & $-$    & $\,\,\,8.1600\pm0.0003$  & \\
HD\,13167\,B        & \,\,\,$6.7931\pm0.1254$   & $\,\,\,\,44.134\pm0.215$   &$-39.358\pm0.212$  & $0.444$ & $3.3$ & $17.4513\pm0.0022$ & $0.2431_{-0.0648}^{+0.0597}$ & \texttt{SHC}\\\hline
HR\,858\,A*         & $31.2565\pm0.0700$  & $123.229\pm0.070$  &$\,\,105.788\pm0.151$  & $0.086$ & $3.9$ & $\,\,\,6.2480\pm0.0003$  & $0.1320_{-0.0911}^{+0.1061}$\\
HR\,858\,B          & $32.3014\pm0.1670$  & $137.125\pm0.213$  &$\,\,105.865\pm0.302$  & $0.835$ & $63$ & $16.0464\pm0.0031$ & \\\hline
HD\,18015\,A*       & \,\,\,$8.0490\pm0.0517$   & $\,\,\,\,63.053\pm0.089$   &$\,\,\,-4.359\pm0.082$   & $-$     & $-$    & $\,\,\,7.7219\pm0.0005$  & \\
HD\,18015\,B        & \,\,\,$7.9413\pm0.0415$   & $\,\,\,\,64.638\pm0.071$   &$\,\,\,-4.668\pm0.066$   & $-$     & $-$    & $12.2361\pm0.0008$ & $0.1440_{-0.0325}^{+0.0361}$\\\hline
K2-288\,B*          & $15.2166\pm0.2007$  & $185.476\pm0.708$  &$-74.070\pm0.618$  & $0.766$ & $70$ & $14.5451\pm0.0017$ & \\
K2-288\,A           & $14.2879\pm0.0807$  & $187.057\pm0.151$  &$-69.591\pm0.116$  & $0.418$ & $41$ & $13.3090\pm0.0009$ & $0.5668_{-0.0824}^{+0.0884}$ & \texttt{SHC}\\\hline
HD\,23472\,A*       & $25.5897\pm0.0261$  & $-102.571\pm0.050\,\,\,\,$ &$-43.917\pm0.059$  & $-$     & $-$    & $\,\,\,9.3848\pm0.0002$  & $0.0783_{-0.0703}^{+0.1742}$\\
HD\,23472\,B        & $25.5060\pm0.0732$  & $-103.019\pm0.154\,\,\,\,$ &$-42.771\pm0.169$  & $0.490$ & $16.1$ & $15.8312\pm0.0014$ & \\\hline
HD\,24085\,B*       & $18.1859\pm0.0245$  & $\,\,\,-9.249\pm0.048$   &$-97.358\pm0.050$  & $-$     & $-$    & $\,\,\,7.4250\pm0.0002$  & \\
HD\,24085\,A        & $18.1430\pm0.0226$  & $-10.234\pm0.043$  &$-97.151\pm0.049$  & $-$     & $-$    & $\,\,\,7.2719\pm0.0002$  & $0.6682_{-0.3903}^{+0.5911}$\\\hline
HII\,1348\,A*\,(SB) & \,\,\,$6.9890\pm0.0490$   & $\,\,\,\,21.401\pm0.120$   &$-45.705\pm0.060$  & $-$     & $-$    & $12.2439\pm0.0012$ & \\
HII\,1348\,C        & \,\,\,$6.6456\pm0.1763$   & $\,\,\,\,20.250\pm0.337$   &$-45.292\pm0.235$  & $0.429$ & $4.1$ & $17.0303\pm0.0017$ & $1.1810_{-0.3321}^{+0.3091}$\\
HII\,1348\,D        & \,\,\,$7.8946\pm1.7831$   & $\,\,\,\,23.361\pm4.630$   &$-42.219\pm2.330$  & $2.369$ & $0.9$ & $20.7790\pm0.0151$ & $0.6550_{-0.4581}^{+0.3718}$\\\hline
HATS-57\,A*         & \,\,\,$3.5495\pm0.0392$   & $-12.664\pm0.046$  &$-14.115\pm0.040$  & $-$     & $-$    & $12.1816\pm0.0007$ & $0.0548_{-0.0423}^{+0.1726}$\\
HATS-57\,B          & \,\,\,$3.4904\pm0.1265$   & $-12.064\pm0.174$  &$-14.764\pm0.142$  & $-$     & $-$    & $17.5558\pm0.0012$ & \\\hline
FU\,Tau\,A*         & \,\,\,$7.5981\pm0.1497$   & $\,\,\,\,\,\,\,6.895\pm0.376$    &$-21.026\pm0.202$  & $0.732$ & $83$ & $15.2412\pm0.0024$ & $2.2620_{-0.4841}^{+0.2597}$\\
FU\,Tau\,B          & \,\,\,$7.4909\pm1.2887$   & $\,\,\,\,12.450\pm4.056$   &$-21.761\pm1.903$  & $3.516$ & $4.6$ & $20.4799\pm0.0074$ & \\\hline
DH\,Tau\,A*         & \,\,\,$7.3880\pm0.0693$   & $\,\,\,\,\,\,\,7.065\pm0.117$    &$-20.699\pm0.079$  & $-$     & $-$    & $12.4961\pm0.0090$ & \\
DH\,Tau\,C          & \,\,\,$7.4011\pm0.0520$   & $\,\,\,\,\,\,\,6.899\pm0.113$    &$-21.207\pm0.074$  & $-$     & $-$    & $11.9692\pm0.0013$ & $2.6683_{-0.1714}^{+0.3698}$\\\hline
51\,Eri\,A*         & $33.5770\pm0.1354$  & $\,\,\,\,44.352\pm0.227$   &$-63.833\pm0.178$  & $0.562$ & $190$  & $\,\,\,5.1224\pm0.0017$  & $0.1740_{-0.1123}^{+0.2663}$\\
51\,Eri\,B\,(SB)    & $37.9633\pm0.3662$  & $\,\,\,\,59.587\pm0.717$   &$-52.419\pm0.618$  & $1.958$ & $2030$ & $\,\,\,9.7247\pm0.0011$  & \\\hline
2M\,0441+23\,C*     & \,\,\,$8.3040\pm0.3778$   & $\,\,\,\,\,\,\,8.955\pm0.931$    &$-21.431\pm0.456$  & $1.350$ & $5.7$ & $18.9668\pm0.0068$ & \\
2M\,0441+23\,AB     & \,\,\,$8.0161\pm0.0832$   & $\,\,\,\,\,\,\,8.300\pm0.189$    &$-21.553\pm0.103$  & $0.529$ & $79$ & $13.8267\pm0.0011$ &$1.0577_{-0.4138}^{+0.9503}$\\
\hline\hline
\end{tabular}
\end{center}
\end{small}
\label{Tab_Gaia_AstroPhoto}
\end{table*}
\end{landscape}
\newpage

\setcounter{table}{1}
\begin{landscape}
\begin{table*}
\caption{continued}
\begin{small}
\begin{center}
\begin{tabular}{lcccccccc}
\hline\hline
Name  & $\pi$         & $\mu_{\alpha}cos(\delta)$ & $\mu_{\delta}$ & $epsi$ & $sig$- & G  & $A_{G}$ \\
      & [mas]              & [mas/yr]                     & [mas/yr]       & [mas]  & $epsi$ & [mag] & [mag]   \\
\hline\hline
NGTS-6\,A*&$\,\,\,3.2151\pm0.0148$&$\,\,\,-9.339\pm0.025$&$-21.9950\pm0.026$&$-$&$-$&$13.8175\pm0.0006$&\\
NGTS-6\,B&$\,\,\,3.2231\pm0.0653$&$\,\,\,-9.301\pm0.107$&$-22.1300\pm0.114$&$0.203$&$1.3$&$17.0603\pm0.0009$&$0.3627_{-0.1403}^{+0.1024}$\\\hline
AB\,Dor\,AC*&$65.3199\pm0.1440$&$\,\,\,\,29.150\pm0.251$&$\,\,164.4210\pm0.299$&$0.850$&$317$&$\,\,\,6.6738\pm0.0018$&\\
AB\,Dor\,BD&$67.0283\pm0.0901$&$\,\,\,\,66.366\pm0.155$&$\,\,125.8990\pm0.189$&$0.522$&$111$&$11.3560\pm0.0012$&$1.3528_{-0.4435}^{+0.5192}$\\\hline
HD\,39855\,A*&$42.9636\pm0.0346$&$\,\,\,\,92.854\pm0.046$&$-24.4660\pm0.063$&$-$&$-$&$\,\,\,7.3211\pm0.0002$&\\
HD\,39855\,B&$42.9612\pm0.0369$&$\,\,\,\,96.166\pm0.055$&$-11.8960\pm0.065$&$-$&$-$&$10.0503\pm0.0006$&$0.1920_{-0.1257}^{+0.2268}$\\\hline
NGTS-10\,A*&$\,\,\,3.0798\pm0.2610$&$\,\,\,-2.323\pm0.343$&$\,\,\,\,10.5270\pm0.395$&$2.152$&$1100$&$14.2604\pm0.0034$&\\
NGTS-10\,B&$\,\,\,0.2965\pm0.0802$&$\,\,\,-1.120\pm0.219$&$\,\,\,\,\,\,\,9.6710\pm0.161$&$0.064$&$0.3$&$15.5926\pm0.0014$&$0.7679_{-0.5770}^{+1.1986}$ & \texttt{SHC}\\\hline
L2\,Pup\,A*&$15.61\pm0.99$&$\,106.31\pm0.96$&$\,\,\,\,324.99\pm1.08$&$-$&$-$&$9.8208\pm0.2812$&&\texttt{HIP}\\
L2\,Pup\,B&$16.4131\pm0.0574$&$\,105.895\pm0.097$&$\,\,327.272\pm0.099$&$0.368$&$12$&$15.7099\pm0.0009$&$0.2940_{-0.0934}^{+0.3029}$\\\hline
HIP\,38594\,A*&$56.1868\pm0.0297$&$-300.905\pm0.044\,\,\,$&$\,\,200.923\pm0.045$&$-$&$-$&$\,\,\,9.0853\pm0.0003$&\\
HIP\,38594\,B&$56.1234\pm0.0799$&$-297.867\pm0.126\,\,\,$&$\,\,206.598\pm0.257$&$-$&$-$&$16.0444\pm0.0005$&$0.3670_{-0.2630}^{+0.7758}$\\\hline
WASP-180\,A*&$\,\,\,3.9093\pm0.0517$&$-14.052\pm0.091$&$\,\,\,-3.169\pm0.055$&$-$&$-$&$10.9134\pm0.0007$&\\
WASP-180\,B&$\,\,\,3.8618\pm0.0734$&$-12.705\pm0.172$&$\,\,\,-2.710\pm0.103$&$-$&$-$&$11.7712\pm0.0008$&$0.0930_{-0.0534}^{+0.1560}$\\\hline
HD\,79211\,B*&$157.8851\pm0.0414\,\,\,$&$-1573.120\pm0.061\,\,\,\,\,\,$&$-660.121\pm0.058\,\,\,\,$&$-$&$-$&$\,\,\,7.0477\pm0.0004$&\\
HD\,79211\,A&$157.8796\pm0.0366\,\,\,$&$-1546.100\pm0.059$\,\,\,\,\,\,&$-569.127\pm0.060\,\,\,\,$&$-$&$-$&$\,\,\,6.9689\pm0.0005$&$0.3757_{-0.2078}^{+0.3264}$\\\hline
HD\,85628\,A*&$\,\,\,5.8297\pm0.0318$&$\,\,\,\,\,\,\,6.051\pm0.055$&$-15.398\pm0.051$&$-$&$-$&$\,\,\,8.1740\pm0.0004$&$0.5012_{-0.2262}^{+0.1438}$\\
HD\,85628\,B&$\,\,\,5.9508\pm0.0366$&$\,\,\,\,\,\,\,5.856\pm0.066$&$-13.252\pm0.060$&$0.296$&$18$&$14.0490\pm0.0047$&\\\hline
TOI\,717\,A*&$28.7709\pm0.0783$&$-26.092\pm0.176$&$\,\,\,\,62.064\pm0.260$&$0.045$&$0.5$&$12.6410\pm0.0005$&$0.3110_{-0.0211}^{+0.1274}$\\
TOI\,717\,B&$28.7588\pm0.0824$&$-23.995\pm0.177$&$\,\,\,\,62.081\pm0.273$&$0.060$&$1.0$&$12.7386\pm0.0008$&\\\hline
G\,196-3\,A*&$45.8611\pm0.0388$&$-141.177\pm0.055\,\,\,$&$-202.394\pm0.053\,\,\,\,$&$0.000$&$0.5$&$10.6123\pm0.0005$&$1.0645_{-0.4836}^{+0.4919}$\\
G\,196-3\,B&$44.3549\pm0.8128$&$-137.820\pm0.928\,\,\,$&$-208.523\pm1.671\,\,\,\,$&$2.210$&$3.3$&$20.1697\pm0.0085$&\\\hline
LTT\,3780\,A*&$45.4644\pm0.0827$&$-341.409\pm0.114\,\,\,$&$-247.870\pm0.105\,\,\,\,$&$0.137$&$6.6$&$11.8465\pm0.0005$&$0.5015_{-0.3436}^{+0.2595}$\\
LTT\,3780\,B&$45.2879\pm0.1081$&$-341.379\pm0.149\,\,\,$&$-248.419\pm0.135\,\,\,\,$&$0.414$&$23$&$14.4855\pm0.0008$&\\\hline
MASCARA-3\,\,A*&$10.3320\pm0.0333$&$-56.184\pm0.053$&$-34.808\pm0.064$&$-$&$-$&$\,\,\,8.2375\pm0.0004$&$0.2400_{-0.1150}^{+0.2740}$\\
MASCARA-3\,\,B&$11.0260\pm0.1268$&$-50.757\pm0.325$&$-37.811\pm0.200$&$0.799$&$160$&$13.0002\pm0.0109$&\\\hline
2M\,J1101-7732\,A*&$\,\,\,5.4081\pm0.1877$&$-22.653\pm0.435$&$\,\,\,\,\,\,\,2.062\pm0.397$&$1.271$&$12$&$18.3299\pm0.0023$&\\
2M\,J1101-7732\,B&$\,\,\,5.4333\pm0.3368$&$-23.668\pm0.748$&$\,\,\,\,\,\,\,1.931\pm0.723$&$1.836$&$6.1$&$19.4040\pm0.0047$&$0.5900_{-0.1200}^{+0.1200}$ & \maltese\\\hline
WASP-175\,A*&$\,\,\,1.8260\pm0.0399$&$-24.306\pm0.057$&$\,\,\,\,\,\,\,6.033\pm0.057$&$-$&$-$&$12.7065\pm0.0002$&\\
WASP-175\,B&$\,\,\,1.7947\pm0.0308$&$-24.064\pm0.045$&$\,\,\,\,\,\,\,6.185\pm0.045$&$-$&$-$&$14.2462\pm0.0003$&$0.0965_{-0.0876}^{+0.0966}$\\
\hline\hline
\end{tabular}
\end{center}
\end{small}
\end{table*}
\end{landscape}
\newpage

\setcounter{table}{1}
\begin{landscape}
\begin{table*}
\caption{continued}
\begin{small}
\begin{center}
\begin{tabular}{lcccccccc}
\hline\hline
Name  & $\pi$         & $\mu_{\alpha}cos(\delta)$ & $\mu_{\delta}$ & $epsi$ & $sig$- & G  & $A_{G}$ \\
           & [mas]              & [mas/yr]                     & [mas/yr]       & [mas]  & $epsi$ & [mag] & [mag]   \\
\hline\hline
CHXR\,73\,A*&$\,\,\,5.2343\pm0.1759$&$-22.193\pm0.233$&$\,\,\,\,\,\,\,0.215\pm0.206$&$0.815$&$12$&$17.2934\pm0.0014$&$3.4650_{-1.0250}^{+1.0250}$ & $\maltese$\\
CHXR\,73\,C&$\,\,\,5.2502\pm0.2218$&$-22.937\pm0.433$&$\,\,\,-1.261\pm0.347$&$1.241$&$14$&$17.9098\pm0.0021$&\\\hline
GJ\,414\,A*&$84.0803\pm0.0471$&$\,\,591.622\pm0.081$&$-197.247\pm0.091\,\,\,\,$&$-$&$-$&$\,\,\,7.7281\pm0.0007$&\\
GJ\,414\,B&$84.1971\pm0.0579$&$\,\,604.831\pm0.081$&$-206.442\pm0.075\,\,\,\,$&$-$&$-$&$\,\,\,9.0471\pm0.0011$&$0.6100_{-0.3221}^{+0.5656}$\\\hline
HD\,97334\,A*&$44.1428\pm0.0383$&$-249.387\pm0.090\,\,\,$&$-151.590\pm0.071\,\,\,\,$&$-$&$-$&$\,\,\,6.2410\pm0.0006$&$0.0555_{-0.0415}^{+0.1848}$\\
HD\,97334\,BC&$42.8724\pm1.1025$&$-236.349\pm2.133\,\,\,$&$-152.068\pm2.109\,\,\,\,$&$5.342$&$15$&$19.9859\pm0.0135$&\\\hline
HD\,233832\,A*&$16.9952\pm0.0752$&$-473.960\pm0.075\,\,\,$&$124.167\pm0.087$&$-$&$-$&$\,\,\,9.9456\pm0.0005$&\\
HD\,233832\,B&$17.0667\pm0.0532$&$-478.645\pm0.048\,\,\,$&$119.122\pm0.088$&$-$&$-$&$12.7187\pm0.0004$&$0.2288_{-0.1079}^{+0.2054}$\\\hline
2M\,J1155-7919\,A*&$\,\,\,9.8862\pm0.0585$&$-41.179\pm0.127$&$\,\,\,-4.336\pm0.086$&$0.490$&$35$&$14.8180\pm0.0017$&$0.4158_{-0.0052}^{+0.0616}$ & $\maltese$\\
2M\,J1155-7919\,B&$\,\,\,9.8211\pm0.5264$&$-39.738\pm1.216$&$\,\,\,-4.656\pm0.687$&$1.387$&$1.4$&$19.9246\pm0.0079$&\\\hline
NGTS-5\,A*&$\,\,\,3.2310\pm0.0272$&$\,\,\,\,13.650\pm0.041$&$\,\,\,-4.688\pm0.042$&$-$&$-$&$13.5260\pm0.0004$&$0.1970_{-0.1174}^{+0.1050}$\\
NGTS-5\,B&$\,\,\,2.9428\pm0.1254$&$\,\,\,\,13.938\pm0.209$&$\,\,\,-4.331\pm0.180$&$0.422$&$3.0$&$17.3160\pm0.0016$&\\\hline
2M\,J1450-7841\,A*&$10.9480\pm0.5046$&$-37.597\pm0.881$&$-23.654\pm0.895$&$2.345$&$5.0$&$19.6858\pm0.0060$&$0.5000_{-0.5000}^{+0.5000}$ & \maltese\\
2M\,J1450-7841\,B&$\,\,\,8.6592\pm0.9023$&$-34.984\pm2.047$&$-22.162\pm1.764$&$2.067$&$1.1$&$20.6501\pm0.0097$&\\\hline
WASP-189\,A*&$\,\,\,9.9990\pm0.0747$&$-50.564\pm0.109$&$-23.788\pm0.115$&$0.082$&$2.2$&$\,\,\,6.5537\pm0.0004$&$0.2652_{-0.2306}^{+0.1599}$ & \texttt{SHC}\\
WASP-189\,B&$10.7202\pm0.1648$&$-50.594\pm0.165$&$-24.037\pm0.178$&$0.475$&$25$&$14.3874\pm0.0024$&\\\hline
HIP\,73990\,A*&$\,\,\,9.0326\pm0.0648$&$-27.432\pm0.106$&$-29.028\pm0.089$&$-$&$-$&$\,\,\,8.0678\pm0.0009$&\\
HIP\,73990\,D&$\,\,\,8.9507\pm0.0899$&$-27.728\pm0.159$&$-29.245\pm0.135$&$0.457$&$28$&$14.6580\pm0.0007$&$1.1073_{-0.4489}^{+0.6427}$\\\hline
TOI\,905\,A*&$\,\,\,6.2745\pm0.0285$&$-25.839\pm0.033$&$-41.150\pm0.051$&$-$&$-$&$11.0813\pm0.0004$&$0.2703_{-0.1606}^{+0.1330}$& \texttt{SHC}\\
TOI\,905\,B&$\,\,\,7.8542\pm0.5489$&$-18.290\pm0.763$&$-39.819\pm0.788$&$1.423$&$25$&$17.2149\pm0.0375$&\\\hline
2M\,1510\,A*&$27.2203\pm0.2665$&$-118.747\pm0.492\,\,\,$&$-46.865\pm0.420$&$1.112$&$18$&$17.4870\pm0.0018$&$0.9266_{-0.0121}^{+0.1360}$ & \texttt{SHC}\\
2M\,1510\,B&$27.6869\pm0.4939$&$-117.448\pm0.893\,\,\,$&$-45.713\pm0.746$&$1.710$&$7.8$&$18.8855\pm0.0035$&\\\hline
$\beta$\,Cir\,A*&$35.1736\pm0.4253$&$-96.742\pm0.491$&$-136.541\pm0.621\,\,\,\,$&$1.852$&$1770$&$\,\,\,3.9732\pm0.0026$&$0.2560_{-0.1598}^{+0.2260}$\\
$\beta$\,Cir\,B&$34.7836\pm0.6840$&$-92.763\pm0.829$&$-138.156\pm1.469\,\,\,\,$&$2.701$&$15$&$19.4335\pm0.0051$&\\\hline
KELT-23\,A*&$\,\,\,7.8912\pm0.0219$&$\,\,\,\,\,\,\,0.434\pm0.039$&$-12.217\pm0.041$&$-$&$-$&$10.1820\pm0.0004$&$0.0680_{-0.0505}^{+0.0996}$\\
KELT-23\,B&$\,\,\,7.8949\pm0.0529$&$\,\,\,\,\,\,\,1.567\pm0.093$&$-11.903\pm0.107$&$0.284$&$6.8$&$15.5209\pm0.0014$&\\\hline
K2-290\,A*B&$\,\,\,3.6365\pm0.0503$&$\,\,\,\,27.225\pm0.099$&$-16.893\pm0.066$&$-$&$-$&$10.8204\pm0.0004$&$0.8900_{-0.2655}^{+0.2570}$\\
K2-290\,C&$\,\,\,4.0531\pm0.2711$&$\,\,\,\,27.465\pm0.593$&$-16.484\pm0.370$&$0.556$&$1.2$&$18.5920\pm0.0027$&\\\hline
GQ\,Lup\,A*&$\,\,\,6.5868\pm0.0473$&$-14.257\pm0.097$&$-23.596\pm0.066$&$0.110$&$5.7$&$11.2608\pm0.0089$&$2.7645_{-0.4140}^{+0.3426}$\\
GQ\,Lup\,C&$\,\,\,5.4925\pm0.4597$&$-14.807\pm0.972$&$-21.947\pm0.653$&$2.960$&$59$&$18.3740\pm0.0037$&\\
\hline\hline
\end{tabular}
\end{center}
\end{small}
\end{table*}
\end{landscape}
\newpage

\setcounter{table}{1}
\begin{landscape}
\begin{table*}
\caption{continued}
\begin{small}
\begin{center}
\begin{tabular}{lcccccccc}
\hline\hline
Name  & $\pi$         & $\mu_{\alpha}cos(\delta)$ & $\mu_{\delta}$ & $epsi$ & $sig$- & G  & $A_{G}$ \\
           & [mas]              & [mas/yr]                     & [mas/yr]       & [mas]  & $epsi$ & [mag] & [mag]   \\
\hline\hline
HIP\,77900\,A*&$\,\,\,6.6037\pm0.1196$&$-13.357\pm0.187$&$-25.272\pm0.110$&$0.212$&$21$&$\,\,\,6.1129\pm0.0006$&$0.1852_{-0.0360}^{+0.1535}$\\
HIP\,77900\,B&$\,\,\,5.2279\pm0.9696$&$-13.908\pm1.517$&$-23.265\pm1.097$&$2.007$&$4.3$&$19.5660\pm0.0057$&\\\hline
USco\,1602-2401\,A*&$\,\,\,6.9484\pm0.0661$&$-11.850\pm0.119$&$-24.032\pm0.051$&$-$&$-$&$11.8656\pm0.0026$&$1.8885_{-0.1186}^{+0.1123}$ & \texttt{SHC}\\
USco\,1602-2401\,B&$\,\,\,6.3381\pm0.2030$&$-12.699\pm0.325$&$-23.872\pm0.187$&$0.829$&$40$&$16.3640\pm0.0010$&\\\hline
HIP\,79098\,A\,(SB)*&$\,\,\,6.8337\pm0.1176$&$\,\,\,-9.823\pm0.210$&$-28.119\pm0.163$&$0.289$&$35$&$\,\,\,5.8264\pm0.0006$&$0.4640_{-0.2381}^{+0.1440}$\\
HIP\,79098\,C&$\,\,\,7.1870\pm0.1476$&$-10.979\pm0.271$&$-26.128\pm0.189$&$0.666$&$27$&$16.1091\pm0.0012$&\\\hline
USco\,1610-1913\,A*&$\,\,\,7.4960\pm0.0718$&$\,\,\,-9.342\pm0.206$&$-23.591\pm0.111$&$0.168$&$12$&$12.6962\pm0.0049$&$0.3850_{-0.3850}^{+0.3850}$ & $\maltese$\\
USco\,1610-1913\,B&$\,\,\,6.9600\pm0.3719$&$\,\,\,-7.043\pm1.112$&$-24.982\pm0.576$&$1.301$&$7.0$&$18.7225\pm0.0029$&\\\hline
USco\,1612-1800\,A*&$\,\,\,6.3156\pm0.0747$&$\,\,\,-7.418\pm0.161$&$-21.148\pm0.112$&$0.427$&$31$&$14.5532\pm0.0007$&$0.3850_{-0.3850}^{+0.3850}$ & $\maltese$\\
USco\,1612-1800\,B&$\,\,\,6.0413\pm0.3224$&$\,\,\,-7.002\pm0.698$&$-19.738\pm0.512$&$1.112$&$4.0$&$18.8810\pm0.0032$&\\\hline
ROXs\,12\,A*&$\,\,\,7.2894\pm0.0417$&$\,\,\,-7.185\pm0.090$&$-24.851\pm0.059$&$0.158$&$6.9$&$13.2655\pm0.0013$&$1.8000_{-1.0000}^{+1.0000}$ & $\maltese$\\
ROXs\,12\,C&$\,\,\,7.2328\pm0.0738$&$\,\,\,-6.577\pm0.151$&$-25.106\pm0.099$&$0.354$&$20$&$14.7659\pm0.0113$&\\\hline
HATS-48\,A*&$\,\,\,3.7648\pm0.0237$&$\,\,\,\,\,\,\,3.125\pm0.031$&$\,\,\,\,\,\,\,6.146\pm0.029$&$-$&$-$&$13.8951\pm0.0002$&$0.5090_{-0.3705}^{+0.1876}$\\
HATS-48\,B&$\,\,\,3.6354\pm0.4432$&$\,\,\,\,\,\,\,2.951\pm0.469$&$\,\,\,\,\,\,\,5.201\pm0.446$&$0.576$&$0.6$&$19.3368\pm0.0037$&\\\hline
GJ\,752\,\,A*&$169.1590\pm0.0520\,\,\,$&$-579.043\pm0.088\,\,\,$&$-1332.740\pm0.081\,\,\,\,\,\,\,$&$-$&$-$&$\,\,\,8.0976\pm0.0011$&\\
GJ\,752\,\,B&$168.9620\pm0.1299\,\,\,$&$-598.177\pm0.245\,\,\,$&$-1365.270\pm0.227\,\,\,\,\,\,$&$0.855$&$98$&$14.3212\pm0.0007$&$1.4208_{-0.1719}^{+0.2795}$\\\hline
HD\,181234\,A*&$20.9155\pm0.0564$&$-122.751\pm0.098\,\,\,$&$-318.277\pm0.098\,\,\,\,$&$-$&$-$&$\,\,\,8.3693\pm0.0004$&$0.3670_{-0.0723}^{+0.1987}$\\
HD\,181234\,B&$20.8683\pm0.1458$&$-117.558\pm0.192\,\,\,$&$-323.292\pm0.211\,\,\,\,$&$0.520$&$27$&$14.2207\pm0.0012$&\\\hline
Wendelstein-1\,A*&$\,\,\,3.2470\pm0.0317$&$\,\,\,\,\,\,\,4.131\pm0.041$&$\,\,\,-1.832\pm0.039$&$-$&$-$&$15.0324\pm0.0005$&$0.5240_{-0.4315}^{+0.4730}$\\
Wendelstein-1\,B&$\,\,\,3.5833\pm0.4053$&$\,\,\,\,\,\,\,3.862\pm0.508$&$\,\,\,-2.194\pm0.536$&$1.038$&$1.9$&$19.3979\pm0.0031$&\\\hline
2M\,J2126-81\,A*&$29.2836\pm0.0690$&$\,\,\,\,59.843\pm0.111$&$-107.723\pm0.114\,\,\,\,$&$0.303$&$48$&$10.8133\pm0.0021$&$0.2157_{-0.0470}^{+0.4319}$ & \texttt{SHC}\\
2M\,J2126-81\,B&$29.2463\pm0.9205$&$\,\,\,\,56.511\pm1.656$&$-115.369\pm2.441$\,\,\,\,&$4.299$&$4.4$&$20.7247\pm0.0094$&\\\hline
TOI\,132\,A*&$\,\,\,6.0809\pm0.0366$&$\,\,\,\,35.553\pm0.043$&$-53.055\pm0.054$&$0.090$&$3.6$&$11.3208\pm0.0007$&$0.1535_{-0.1430}^{+0.1606}$\\
TOI\,132\,B&$\,\,\,5.9683\pm0.2251$&$\,\,\,\,35.417\pm0.280$&$-52.488\pm0.361$&$0.955$&$7.4$&$18.4470\pm0.0015$&\\\hline
NGTS-7\,A*&$\,\,\,7.2497\pm0.1203$&$-27.003\pm0.114$&$-16.225\pm0.178$&$0.610$&$72$&$14.9154\pm0.0020$&$0.5000_{-0.5000}^{+0.5000}$ & $\maltese$\\
NGTS-7\,B&$\,\,\,6.5232\pm0.0787$&$-28.601\pm0.112$&$-14.776\pm0.364$&$0.203$&$4.4$&$15.5134\pm0.0012$&\\\hline
DS\,Tuc\,A*&$22.6663\pm0.0354$&$\,\,\,\,79.464\pm0.074$&$-67.440\pm0.045$&$-$&$-$&$\,\,\,8.3193\pm0.0010$&\\
DS\,Tuc\,B&$22.6504\pm0.0297$&$\,\,\,\,78.022\pm0.064$&$-65.746\pm0.037$&$-$&$-$&$\,\,\,9.3993\pm0.0014$&$0.3210_{-0.1034}^{+0.2350}$\\\hline
1RXS\,J2351+3127\,A*&$23.2183\pm0.0524$&$\,\,106.584\pm0.064$&$-87.761\pm0.038$&$0.083$&$3.9$&$12.5145\pm0.0005$&\\
1RXS\,J2351+3127\,C&$23.1794\pm0.0592$&$\,\,105.757\pm0.070$&$-87.787\pm0.041$&$0.285$&$32$&$13.2004\pm0.0006$&$0.4190_{-0.0410}^{+0.4250}$\\
\hline\hline
\end{tabular}
\end{center}
\end{small}
\end{table*}
\end{landscape}
\newpage

\textbf{\underline{Comments on individual objects:}}

\begin{itemize}

\small

\item \textbf{HD\,1160\,A} hosts a brown dwarf companion \citep[HD\,1160\,B, detected by][]{nielsen2012}, which is listed as exoplanet in the \texttt{EPE}.

\item The exoplanet host star \textbf{HD\,24085\,B} is the secondary component of a binary system, whose primary star HD\,24085\,A is also known as HD\,24062.

\item \textbf{HII\,1348\,A} is a spectroscopic binary with a brown dwarf companion \citep[HII\,1348\,B, discovered by][]{Geissler2012}, which is listed as exoplanet in the \texttt{EPE}.

\item \textbf{DH\,Tau\,A} hosts a brown dwarf companion (DH\,Tau\,B), which was detected by \cite{itoh2005} and is listed as exoplanet in the \texttt{EPE}. DH\,Tau\,C (alias DI\,Tau) is the wide primary component of this system.

\item \textbf{2M\,0441+23\,C} is an exoplanet host brown dwarf \citep{bowler2015}, which is listed in the \texttt{EPE}.

\item The bright AGB star \textbf{L2\,Pup\,A} is listed in the Gaia DR2 but with a parallax ($\pi=7.3644\pm0.6149$\,mas) that significantly differs from its HIPPARCOS-value \citep[$\pi=15.61\pm0.99$\,mas, ][]{vanleeuwen2007}. Furthermore, it should be noted that the G-band brightness of this star, as listed in the Gaia DR2, is several magnitudes fainter than expected \citep[e.g. $G=3.97\pm0.54$\,mag, as estimated by][]{smart2014}. Therefore, we only use here the Gaia DR2 equatorial coordinates of this star, while we adopt the HIPPARCOS-values of its parallax and proper motion, which is indicated with the flag \texttt{HIP} in this table.
\item \textbf{HIP\,73990\,A} is the host star of two brown dwarfs \citep[HIP\,73990\,B \& C, revealed by][]{hinkley2015}, which are both listed as exoplanets in the \texttt{EPE}.

\item \textbf{GQ\,Lup\,A} is listed as exoplanet host star in the \texttt{EPE}, whose substellar companion was detected by \cite{neuhaeuser2005}. The star exhibits a wide stellar companion, whose WDS designation (GQ\,Lup\,C) is used here.

\item \textbf{HIP\,79098\,A} is a spectroscopic binary and hosts the brown dwarf HIP\,79098\,B \citep{janson2019}, which is listed as exoplanet in the \texttt{EPE}.

\item \textbf{ROXs\,12\,A} is the host star of the brown dwarf ROXs\,12\,B, detected by \cite{kraus2014}, which is listed as exoplanet in the \texttt{EPE}.

\item \textbf{RXSJ2351+3127\,A} hosts a brown dwarf companion \citep[RXSJ2351+3127\,B, discovered by][]{bowler2012}, which is listed as exoplanet in the \texttt{EPE}.

\end{itemize}

\rule{17.4cm}{1pt}

\begin{itemize}

\item \textbf{HII\,1348\,A}, \textbf{FU\,Tau\,A}, \textbf{G\,196-3\,A}, \textbf{2M\,J1155-7919\,A}, \textbf{HD\,97334\,A}, \textbf{$\beta$\,Cir\,A}, \textbf{HIP\,77900\,A}, \textbf{USco\,1602-2401\,A}, \textbf{USco\,1610-1913\,A}, \textbf{USco\,1612-1800\,A}, and \textbf{2M\,J2126-81\,A}, are all listed as exoplanet host stars in the \texttt{EPE}, whose substellar companions were detected and characterized in this study, using data from the Gaia DR2.

\item \textbf{2M\,J1101-7732\,A}, \textbf{2M\,J1450-7841\,A}, \textbf{2M\,1510\,A} are all brown dwarfs, which are listed as exoplanet hosts in the \texttt{EPE}, whose substellar companions were detected and characterized in this study with Gaia DR2 data.

\end{itemize}

\newpage

\begin{landscape}
\begin{table*} \caption{The relative astrometry and WDS status of all detected companions.}
\begin{tabular}{lcccccccc}
\hline\hline
Companion    & $\rho$    & $PA$        & $\Delta\pi$ & $sig$-      & $\mu_{rel}$ & $sig$-      & $cpm$- & not in\\
             & [arcsec]  & [$^\circ$]  & [mas]       & $\Delta\pi$ & [mas/yr]    & $\mu_{rel}$ & $index$ & WDS\\
\hline\hline
HD\,1160\,C&$5.14549\pm0.00018$&$349.53223\pm0.00259$&$0.95\pm0.28$&$3.3~(1.2)$&$1.72\pm0.33$&$5.2$&$30$&$$\\\hline
Gliese\,49\,B&$294.45989\pm0.00011\,\,\,\,\,\,$&$\,\,\,75.52728\pm0.00002$&$0.17\pm0.09$&$2.0 ~(0.8)$&$4.36\pm0.23$&$19$&$338$&$$\\\hline
HD\,8326\,B&$56.88131\pm0.00005\,\,\,$&$147.16909\pm0.00006$&$0.12\pm0.08$&$1.6 ~(0.3)$&$1.18\pm0.17$&$7.1$&$394$&$$\\\hline
HD\,13167\,B&$20.06421\pm0.00010\,\,\,$&$\,\,\,24.77589\pm0.00028$&$0.11\pm0.13$&$0.8 ~(0.2)$&$1.28\pm0.23$&$5.7$&$91$&$\bigstar~~~~$\\\hline
HR\,858\,B&$8.35742\pm0.00013$&$\,\,\,15.79337\pm0.00060$&$1.04\pm0.18$&$5.8 ~(1.2)$&$13.90\pm0.22\,\,\,$&$62$&$24$&$\bigstar^{1}~~$\\\hline
HD\,18015\,B&$7.08916\pm0.00006$&$316.16832\pm0.00045$&$0.11\pm0.07$&$1.6 ~(1.6)$&$1.61\pm0.11$&$14$&$79$&$$\\\hline
K2-288\,A&$0.78692\pm0.00018$&$340.38240\pm0.01437$&$0.93\pm0.22$&$4.3 ~(1.0)$&$4.75\pm0.64$&$7.4$&$84$&$\bigstar^{2}~~$\\\hline
HD\,23472\,B&$9.56924\pm0.00008$&$\,\,\,45.28294\pm0.00046$&$0.08\pm0.08$&$1.1 ~(0.2)$&$1.23\pm0.18$&$7.0$&$181$&$\bigstar~~~~$\\\hline
HD\,24085\,A&$75.91260\pm0.00003\,\,\,$&$263.05666\pm0.00002$&$0.04\pm0.03$&$1.3 ~(1.3)$&$1.01\pm0.06$&$16$&$194$&$\bigstar~~~~$\\\hline
HII\,1348\,C&$36.02849\pm0.00016\,\,\,$&$276.87735\pm0.00017$&$0.34\pm0.18$&$1.9 ~(0.7)$&$1.22\pm0.35$&$3.5$&$82$&$\bigstar~~~~$\\\hline
HII\,1348\,D&$55.01726\pm0.00090\,\,\,$&$182.06892\pm0.00182$&$0.91\pm1.78$&$0.5 ~(0.3)$&$4.00\pm3.05$&$1.3$&$25$&$\bigstar~~~~$\\\hline
HATS-57\,B&$14.44086\pm0.00010\,\,\,$&$282.25351\pm0.00032$&$0.06\pm0.13$&$0.4 ~(0.4)$&$0.88\pm0.16$&$5.4$&$43$&$\bigstar~~~~$\\\hline
FU\,Tau\,B&$5.68952\pm0.00112$&$123.57637\pm0.00858$&$0.11\pm1.30$&$0.1 ~(0.0)$&$5.60\pm4.05$&$1.4$&$8$&$$\\\hline
DH\,Tau\,C&$15.29981\pm0.00007\,\,\,$&$126.08805\pm0.00023$&$0.01\pm0.09$&$0.2 ~(0.2)$&$0.53\pm0.11$&$4.7$&$83$&$$\\\hline
51\,Eri\,B\,~(SB)&$66.96749\pm0.00027\,\,\,$&$162.62918\pm0.00028$&$4.39\pm0.39$&$11.2 ~(2.1)\,\,$&$19.04\pm0.71\,\,\,$&$27$&$8$&$$\\\hline
2M\,0441+23\,AB&$12.31449\pm0.00036\,\,\,$&$\,\,\,57.55273\pm0.00133$&$0.29\pm0.39$&$0.7 ~(0.2)$&$0.67\pm0.94$&$0.7$&$70$&$$\\\hline
NGTS-6\,B&$5.36108\pm0.00005$&$116.68846\pm0.00060$&$0.01\pm0.07$&$0.1 ~(0.0)$&$0.14\pm0.12$&$1.2$&$342$&$\bigstar~~~~$\\\hline
AB\,Dor\,BD&$8.87930\pm0.00018$&$347.19358\pm0.00097$&$1.71\pm0.17$&$10.1 ~(1.7)\,\,$&$53.56\pm0.33\,\,\,$&$164$&$6$&$$\\\hline
HD\,39855\,B&$10.72622\pm0.00004\,\,\,$&$\,\,\,19.55064\pm0.00017$&$0.00\pm0.05$&$0.0 ~(0.0)$&$13.00\pm0.09\,\,\,$&$145$&$15$&$$\\\hline
NGTS-10\,B&$1.12234\pm0.00023$&$334.73644\pm0.01107$&$2.78\pm0.27$&$10.2 ~(1.3)\,\,$&$1.48\pm0.41$&$3.6$&$14$&$\bigstar^{3}~~$\\\hline
L2\,Pup\,B&$32.80132\pm0.00052\,\,\,$&$\,\,\,63.66528\pm0.00099$&$0.80\pm0.99$&$0.8\,\,\, ~(-)$&$2.32\pm1.08$&$2.2$&$296$&$\bigstar~~~~$\\\hline
HIP\,38594\,B&$399.81589\pm0.00012\,\,\,\,\,\,$&$208.91546\pm0.00001$&$0.06\pm0.09$&$0.7 ~(0.7)$&$6.44\pm0.24$&$27$&$113$&$$\\\hline
WASP-180\,B&$4.86185\pm0.00006$&$138.92126\pm0.00081$&$0.05\pm0.09$&$0.5 ~(0.5)$&$1.42\pm0.19$&$7.6$&$19$&$$\\\hline
HD\,79211\,A&$17.08255\pm0.00004\,\,\,$&$277.72812\pm0.00014$&$0.01\pm0.06$&$0.1 ~(0.1)$&$94.92\pm0.08\,\,\,$&$1136$&$35$&$$\\\hline
HD\,85628\,B&$4.33622\pm0.00004$&$224.93946\pm0.00056$&$0.12\pm0.05$&$2.5 ~(0.4)$&$2.15\pm0.08$&$27$&$14$&$\bigstar^{4}~~$\\\hline
TOI\,717\,B&$65.46692\pm0.00016\,\,\,$&$\,\,\,88.63021\pm0.00020$&$0.01\pm0.11$&$0.1 ~(0.1)$&$2.10\pm0.25$&$8.4$&$64$&$$\\\hline
G\,196-3\,B&$16.06941\pm0.00055\,\,\,$&$209.15563\pm0.00166$&$1.51\pm0.81$&$1.9 ~(0.6)$&$6.99\pm1.53$&$4.6$&$71$&$$\\\hline
LTT\,3780\,B&$15.78849\pm0.00011\,\,\,$&$\,\,\,97.14133\pm0.00038$&$0.18\pm0.14$&$1.3 ~(0.4)$&$0.55\pm0.17$&$3.2$&$1535$&$$\\\hline
MASCARA-3\,\,B&$2.06449\pm0.00010$&$173.15273\pm0.00345$&$0.69\pm0.13$&$5.3 ~(0.9)$&$6.20\pm0.31$&$20$&$21$&$\bigstar^{5}~~$\\\hline
2M\,J1101-7732\,B&$1.42656\pm0.00041$&$\,\,\,30.00553\pm0.01621$&$0.03\pm0.39$&$0.1 ~(0.0)$&$1.02\pm0.86$&$1.2$&$45$&$$\\\hline
WASP-175\,B&$7.25020\pm0.00003$&$\,\,\,\,\,\,4.95541\pm0.00027$&$0.03\pm0.05$&$0.6 ~(0.6)$&$0.29\pm0.07$&$3.9$&$175$&$\bigstar^{6}~~$\\\hline\hline
\end{tabular}
\label{Tab_CompAstro}
\end{table*}
\end{landscape}
\newpage

\setcounter{table}{2}
\begin{landscape}
\begin{table*}
\caption{continued}
\begin{tabular}{lcccccccc}
\hline\hline
Companion    & $\rho$    & $PA$        & $\Delta\pi$ & $sig$-      & $\mu_{rel}$ & $sig$-      & $cpm$- & not in\\
             & [arcsec]  & [$^\circ$]  & [mas]       & $\Delta\pi$ & [mas/yr]    & $\mu_{rel}$ & $index$ & WDS\\
\hline\hline
CHXR\,73\,C&$46.10344\pm0.00027\,\,\,$&$248.30821\pm0.00031$&$0.02\pm0.28$&$0.1 ~(0.0)$&$1.65\pm0.42$&$3.9$&$27$&$\bigstar~~~~$\\\hline
GJ\,414\,B&$34.15873\pm0.00007\,\,\,$&$262.44625\pm0.00011$&$0.12\pm0.07$&$1.6 ~(1.6)$&$16.09\pm0.12\,\,\,$&$139$&$78$&$$\\\hline
HD\,97334\,BC&$89.88421\pm0.00098\,\,\,$&$245.04583\pm0.00060$&$1.27\pm1.10$&$1.2 ~(0.2)$&$13.05\pm2.13\,\,\,$&$6.1$&$44$&$$\\\hline
HD\,233832\,B&$4.93691\pm0.00004$&$266.38672\pm0.00079$&$0.07\pm0.09$&$0.8 ~(0.8)$&$6.88\pm0.11$&$63$&$143$&$$\\\hline
2M\,J1155-7919\,B&$5.75435\pm0.00047$&$227.86140\pm0.00458$&$0.07\pm0.53$&$0.1 ~(0.0)$&$1.48\pm1.20$&$1.2$&$55$&$\bigstar^{7}~~$\\\hline
NGTS-5\,B&$26.89147\pm0.00011\,\,\,$&$116.31597\pm0.00021$&$0.29\pm0.13$&$2.2 ~(0.7)$&$0.46\pm0.20$&$2.3$&$63$&$\bigstar~~~~$\\\hline
2M\,J1450-7841\,B&$4.23901\pm0.00099$&$313.26065\pm0.01318$&$2.29\pm1.03$&$2.2 ~(0.7)$&$3.01\pm2.17$&$1.4$&$29$&$\bigstar^{8}~~$\\\hline
WASP-189\,B&$9.41610\pm0.00010$&$\,\,\,70.78901\pm0.00095$&$0.72\pm0.18$&$4.0 ~(1.4)$&$0.25\pm0.21$&$1.2$&$446$&$\bigstar~~~~$\\\hline
HIP\,73990\,D&$47.27427\pm0.00009\,\,\,$&$\,\,\,56.65125\pm0.00010$&$0.08\pm0.11$&$0.7 ~(0.2)$&$0.37\pm0.18$&$2.0$&$219$&$\bigstar~~~~$\\\hline
TOI\,905\,B&$2.24803\pm0.00050$&$100.34253\pm0.01658$&$1.58\pm0.55$&$2.9 ~(1.0)$&$7.67\pm0.76$&$10$&$12$&$$\\\hline
2M\,1510\,B&$6.77139\pm0.00046$&$209.28499\pm0.00431$&$0.47\pm0.56$&$0.8 ~(0.2)$&$1.74\pm0.95$&$1.8$&$146$&$\bigstar^{9}~~$\\\hline
$\beta$\,Cir\,B&$217.62247\pm0.00055\,\,\,\,\,\,$&$199.25875\pm0.00013$&$0.39\pm0.81$&$0.5 ~(0.1)$&$4.29\pm1.08$&$4.0$&$78$&$$\\\hline
KELT-23\,B&$4.54135\pm0.00006$&$127.68919\pm0.00069$&$0.00\pm0.06$&$0.1 ~(0.0)$&$1.18\pm0.10$&$12$&$21$&$\bigstar^{10}$\\\hline
K2-290\,C&$11.25119\pm0.00017\,\,\,$&$179.97609\pm0.00151$&$0.42\pm0.28$&$1.5 ~(0.7)$&$0.47\pm0.44$&$1.1$&$135$&$\bigstar^{11}$\\\hline
GQ\,Lup\,C&$16.11286\pm0.00039\,\,\,$&$114.61327\pm0.00099$&$1.09\pm0.46$&$2.4 ~(0.4)$&$1.74\pm0.70$&$2.5$&$31$&$$\\\hline
HIP\,77900\,B&$22.27990\pm0.00044\,\,\,$&$\,\,\,12.74996\pm0.00267$&$1.38\pm0.98$&$1.4 ~(0.6)$&$2.08\pm1.14$&$1.8$&$27$&$\bigstar^{12}$\\\hline
USco\,1602-2401\,B&$7.21512\pm0.00008$&$353.20771\pm0.00157$&$0.61\pm0.21$&$2.9 ~(0.7)$&$0.86\pm0.34$&$2.5$&$62$&$$\\\hline
HIP\,79098\,C&$65.29721\pm0.00018\,\,\,$&$101.86730\pm0.00008$&$0.35\pm0.19$&$1.9 ~(0.5)$&$2.30\pm0.28$&$8.3$&$25$&$\bigstar^{13}$\\\hline
USco\,1610-1913\,B&$5.82725\pm0.00040$&$113.57990\pm0.00238$&$0.54\pm0.38$&$1.4 ~(0.4)$&$2.69\pm1.01$&$2.7$&$19$&$$\\\hline
USco\,1612-1800\,B&$3.18438\pm0.00019$&$\,\,\,10.65437\pm0.00561$&$0.27\pm0.33$&$0.8 ~(0.2)$&$1.47\pm0.54$&$2.7$&$29$&$\bigstar^{14}$\\\hline
ROXs\,12\,C&$37.14026\pm0.00004\,\,\,$&$185.99483\pm0.00012$&$0.06\pm0.08$&$0.7 ~(0.1)$&$0.66\pm0.17$&$3.9$&$79$&$\bigstar^{15}$\\\hline
HATS-48\,B&$5.43813\pm0.00025$&$267.59024\pm0.00322$&$0.13\pm0.44$&$0.3 ~(0.2)$&$0.96\pm0.45$&$2.2$&$13$&$\bigstar^{16}$\\\hline
GJ\,752\,\,B&$75.48951\pm0.00011\,\,\,$&$152.49075\pm0.00009$&$0.20\pm0.14$&$1.4 ~(0.2)$&$37.74\pm0.25\,\,\,$&$153$&$78$&$$\\\hline
HD\,181234\,B&$5.17023\pm0.00011$&$\,\,\,56.61253\pm0.00118$&$0.05\pm0.16$&$0.3 ~(0.1)$&$7.22\pm0.22$&$32$&$95$&$$\\\hline
Wendelstein-1\,B&$11.79208\pm0.00024\,\,\,$&$232.62838\pm0.00116$&$0.34\pm0.41$&$0.8 ~(0.3)$&$0.45\pm0.53$&$0.9$&$20$&$\bigstar~~~~$\\\hline
2M\,J2126-81\,B&$217.49441\pm0.00082\,\,\,\,\,\,$&$123.98914\pm0.00024$&$0.04\pm0.92$&$0.0 ~(0.0)$&$8.34\pm2.34$&$3.6$&$30$&$$\\\hline
TOI\,132\,B&$19.64887\pm0.00018\,\,\,$&$151.44437\pm0.00044$&$0.11\pm0.23$&$0.5 ~(0.1)$&$0.58\pm0.36$&$1.6$&$218$&$\bigstar~~~~$\\\hline
NGTS-7\,B&$1.13095\pm0.00014$&$117.57142\pm0.01072$&$0.73\pm0.14$&$5.1 ~(1.1)$&$2.16\pm0.30$&$7.3$&$30$&$\bigstar^{17}$\\\hline
DS\,Tuc\,B&$5.36461\pm0.00003$&$347.65815\pm0.00047$&$0.02\pm0.05$&$0.3 ~(0.3)$&$2.22\pm0.08$&$29$&$93$&$$\\\hline
1RXS\,J2351+3127\,C&$126.01641\pm0.00005\,\,\,\,\,\,$&$\,\,\,98.50769\pm0.00002$&$0.04\pm0.08$&$0.5 ~(0.1)$&$0.83\pm0.09$&$8.7$&$333$&$$\\
\hline\hline
\end{tabular}
\end{table*}
\end{landscape}
\newpage

\textbf{\underline{Comments on individual companions:}}

\begin{itemize}

\item[1:] This companion was first reported by \cite{vanderburg2019}, who have already verified its equidistance and common proper motion with the exoplanet host star HR\,858\,A using Gaia DR2 data, consistent with the results, obtained in this study.

\item[2:] This companion was detected by \cite{feinstein2019} and its companionship with the exoplanet host star K2-288\,B was proven with Gaia DR2 astrometry, confirmed by the astrometric analysis, carried out in the study, presented here.

\item[3:] This star was already noticed in the Gaia DR2 by \cite{mccormac2020} as common proper motion companion of the exoplanet host star NGTS-10\,A, consistent with the results, derived in this study.

\item[4:] This companion of the exoplanet host star HD\,85628\,A was discovered by \cite{dorval2020} in the Gaia DR2, who found its parallax and proper motion consistent with that of the exoplanet host star, confirmed by the astrometric analysis, presented here.

\item[5:] This companion was detected with AO imaging by \cite{rodriguez2019} using Keck/NIRC\,2, but is also listed in the Gaia DR2, whose astrometry was used by this team to verify the equidistance and common proper motion of this companion with the exoplanet host star MASCARA-3\,A, as done in this study.

\item[6:] This companion was already reported by \cite{nielsen2019}, who proved its companionship with the exoplanet host star WASP-175\,A with Gaia DR2 astrometry, consistent with the results derived here.

\item[7:] The equidistance and common proper motion of this substellar object with the exoplanet host star 2M\,J1155-7919\,A was verified by \cite{dickson-andervelde2020} using Gaia DR2 data, as done in this work.

\item[8:] This companion was detected by \cite{burgasser2017} and its common proper motion with the brown dwarf 2M\,J1450-7841\,A, listed in the \texttt{EPE}, was verified with ground based astrometry, confirmed in this study with Gaia DR2 data, which furthermore proves the equidistance of both objects.

\item[9:] This companion was noticed by \cite{triaud2020} in the Gaia DR2 as equidistant and co-moving companion of the brown dwarf 2M\,1510\,A, which is listed in the \texttt{EPE}, consistent with our results.

\item[10:] KELT-23\,B was first discovered by \citep{johns2019} with Keck/NIRC\,2 AO imaging, who used Gaia DR2 astrometry to prove the equidistance and common proper motion of the companion with the exoplanet host star KELT-23\,A, as done in this study.

\item[11:] This companion was already described by \cite{hjorth2019}, who have verified it to be equidistant and co-moving with the exoplanet host star K2-290\,A, using Gaia DR2 data, a conclusion, which is confirmed by the analysis, presented here. Furthermore, this team identified an additional but closer stellar companion-candidate of the exoplanet host star (K2-290\,B) with Subrau/IRCS AO imaging, which however still needs astrometric confirmation of its companionship. Due to its close angular separation to K2-290\,A we adopt here this object as companion of the exoplanet host star.

\item[12:] This companion was revealed spectro-photometrically by \cite{aller2013}. With Gaia DR2 astrometry we prove here its companionship with the exoplanet host star HIP\,77900\,A.

\item[13:] HIP\,79098\,C was reported by \citep{janson2019} as equidistant and co-moving companion of the exoplanet host star HIP\,79098\,A, based on its Gaia DR2 astrometry, confirmed by the analysis of the companion, which is presented here.\vspace{14mm}

\end{itemize}

\begin{itemize}

\item[14:] This companion was revealed spectro-photometrically by \cite{aller2013}. The equidistance and common proper motion of this companion with the exoplanet host star USco\,1612-1800\,A was proven in this study, with Gaia DR2 astrometry.

\item[15:] This star was identified by \citep{bowler2017} as companion of ROXs\,12\,A, based on its radial velocity and proper motion. We prove the equidistance of both stars with their Gaia DR2 astrometry, which also confirms their common proper motion.

\item[16:] This companion was reported by \citep{hartman2020}, who used the Gaia DR2 astrometry to confirm its companionship with the exoplanet host star HATS-48\,A, as done in this work.

\item[17:] NGTS-7\,B was revealed by \citep{jackman2019} as companion of the exoplanet host star NGTS-7\,A using Gaia DR2 astrometry, as done in this study.

\end{itemize}

\newpage

\begin{landscape}
\begin{table*} \caption{The equatorial coordinates and derived physical properties of all detected companions.}
\begin{tabular}{lccccccc}
\hline\hline

Companion & $\alpha$   & $\delta$   & $M_{G}$ & $sep$ & $mass$        &  $T_{eff}$ & Flags \\
          & [$^\circ$] & [$^\circ$] & [mag]   & [au]  & [$M_{\odot}$] &  [K]       & \\
\hline\hline
HD\,1160\,C&$\,\,\,\,\,\,3.98858470355$&$\,\,\,\,\,\,4.25245524714$&$\,\,9.72_{-0.13}^{+0.10}$&$648$&$0.378_{-0.016}^{+0.021}$&$3503_{-19}^{+25}$ & \texttt{BPRP PRI}\\\hline
Gliese\,49\,B&$\,\,\,15.83942593995$&$\,\,\,\,62.36588062663$&$11.35_{-0.22}^{+0.41}$&$2902$&$0.169_{-0.030}^{+0.020}$&$3211_{-74}^{+40}$ & \texttt{2MA BPRP PRI}\\\hline
HD\,8326\,B&$\,\,\,20.54104760211$&$-26.90734439784$&$11.48_{-0.25}^{+0.04}$&$1747$&$0.198_{-0.003}^{+0.023}$&$3257_{-7}^{+37}$ & \texttt{2MA BPRP PRI}\\\hline
HD\,13167\,B&$\,\,\,32.06020575541$&$-24.69051459458$&$11.33_{-0.06}^{+0.07}$&$3001$&$0.211_{-0.007}^{+0.006}$&$3279_{-10}^{+9}$ & \texttt{2MA BPRP}\\\hline
HR\,858\,B&$\,\,\,42.98571355919$&$-30.81182706733$&$13.39_{-0.11}^{+0.09}$&$267$&$0.112_{-0.002}^{+0.003}$&$2926_{-19}^{+20}$ & \texttt{BPRP PRI}\\\hline
HD\,18015\,B&$\,\,\,43.36225373946$&$\,\,\,-8.84661863284$&$\,\,6.62_{-0.04}^{+0.04}$&$881$&$0.747_{-0.004}^{+0.004}$&$4650_{-18}^{+20}$ & \texttt{2MA BPRP PRI}\\\hline
K2-288\,A&$\,\,\,55.44420842474$&$\,\,\,\,18.26869720107$&$\,\,8.65_{-0.09}^{+0.09}$&$52$&$0.534_{-0.010}^{+0.010}$&$3777_{-28}^{+30}$ & \texttt{2MA}\\\hline
HD\,23472\,B&$\,\,\,55.46315889099$&$-62.76539647729$&$12.79_{-0.18}^{+0.07}$&$374$&$0.129_{-0.002}^{+0.006}$&$3038_{-13}^{+30}$ & \texttt{2MA BPRP PRI}\\\hline
HD\,24085\,A&$\,\,\,56.19506729271$&$-70.02706843481$&$\,\,2.90_{-0.59}^{+0.39}$&$4174$&$1.254_{-0.058}^{+0.088}$&$\,\,6339_{-139}^{+211}$ & \texttt{2MA BPRP PRI EXT}\\\hline
HII\,1348\,C&$\,\,\,56.81444339267$&$\,\,\,\,24.39176993016$&$10.07_{-0.31}^{+0.33}$&$5155$&$0.322_{-0.046}^{+0.048}$&$3436_{-60}^{+58}$ & \texttt{2MA BPRP PRI ***}\\\hline
HII\,1348\,D&$\,\,\,56.82474725542$&$\,\,\,\,24.37529892928$&$14.35_{-0.37}^{+0.46}$&$7872$&$0.055_{-0.006}^{+0.005}$&$2602_{-96}^{+78}$ & \texttt{BD 2MA BPRP ***}\\\hline
HATS-57\,B&$\,\,\,60.94413061754$&$-19.05596683949$&$10.25_{-0.18}^{+0.05}$&$4068$&$0.331_{-0.007}^{+0.023}$&$3448_{-7}^{+24}$ & \texttt{2MA BPRP}\\\hline
FU\,Tau\,B&$\,\,\,65.89894960541$&$\,\,\,\,25.04979704793$&$12.62_{-0.26}^{+0.49}$&$749$&$0.018_{-0.003}^{+0.002}$&$2553_{-71}^{+38}$ & \texttt{BD 2MA BPRP}\\\hline
DH\,Tau\,C&$\,\,\,67.42700661202$&$\,\,\,\,26.54688636442$&$\,\,3.64_{-0.37}^{+0.17}$&$2071$&$1.655_{-0.106}^{+0.228}$&$\,\,4837_{-77}^{+164}$ & \texttt{2MA BPRP PRI EXT}\\\hline
51\,Eri\,B\,(SB)&$\,\,\,69.40630139702$&$\,\,\,-2.49157819607$&$\,\,7.18_{-0.27}^{+0.11}$&$1994$&$0.733_{-0.024}^{+0.056}$&$\,\,3962_{-49}^{+116}$ & \texttt{2MA BPRP PRI ***}\\\hline
2M\,0441+23\,AB$^1$ &$\,\,\,70.44024465438$&$\,\,\,\,23.03268965375$&$\,\,7.37_{-0.96}^{+0.43}$&$1483$&$0.241_{-0.061}^{+0.147}$&$\,\,3308_{-138}^{+300}$ & \texttt{2MA BPRP}\\\hline
NGTS-6\,B&$\,\,\,75.79692465277$&$-30.40013000839$&$\,\,9.23_{-0.10}^{+0.14}$&$1667$&$0.457_{-0.018}^{+0.013}$&$3618_{-29}^{+21}$ & \texttt{2MA BPRP PRI}\\\hline
AB\,Dor\,BD&$\,\,\,82.18603484814$&$-65.44557858318$&$\,\,9.08_{-0.52}^{+0.44}$&$136$&$0.466_{-0.059}^{+0.061}$&$\,\,3645_{-104}^{+167}$ & \texttt{2MA BPRP PRI ***}\\\hline
HD\,39855\,B&$\,\,\,88.62713212305$&$-19.70163948841$&$\,\,8.02_{-0.23}^{+0.13}$&$250$&$0.598_{-0.014}^{+0.024}$&$\,\,3978_{-41}^{+103}$ & \texttt{2MA BPRP PRI}\\\hline
NGTS-10\,B&$\,\,\,91.87213254706$&$-25.59461438962$&$\,\,7.27_{-1.21}^{+0.61}$&$364$&$0.665_{-0.059}^{+0.121}$&$\,\,4327_{-284}^{+606}$ & \\\hline
L2\,Pup\,B&$108.39678197580$&$-44.63427669967$&$ 11.39_{-0.33}^{+0.17}$&$ 2101 $&$ 0.203_{-0.012}^{+0.032}$ & $3270_{-26}^{+49}$ & \\\hline
HIP\,38594\,B&$118.48449010900$&$-25.39952189079$&$14.43_{-0.78}^{+0.26}$&$7116$&$\sim0.6$&$$ & \texttt{WD 2MA BPRP PRI}\\\hline
WASP-180\,B&$123.39313443835$&$\,\,\,-1.98380547425$&$\,\,4.64_{-0.16}^{+0.06}$&$1244$&$1.057_{-0.011}^{+0.029}$&$5778_{-36}^{+93}$ & \texttt{2MA BPRP PRI}\\\hline
HD\,79211\,A&$138.58391575741$&$\,\,\,\,52.68415915741$&$\,\,7.59_{-0.33}^{+0.21}$&$108$&$0.644_{-0.022}^{+0.034}$&$\,\,4180_{-97}^{+153}$ & \texttt{2MA BPRP PRI}\\
\hline\hline
\end{tabular}
\label{Tab_CompProps}
\end{table*}
\end{landscape}

\setcounter{table}{3}
\begin{landscape}
\begin{table*}
\caption{continued}
\begin{tabular}{lccccccc}
\hline\hline
Companion & $\alpha$   & $\delta$   & $M_{G}$ & $sep$ & $mass$        &  $T_{eff}$ & Flags \\
          & [$^\circ$] & [$^\circ$] & [mag]   & [au]  & [$M_{\odot}$] &  [K]       & \\
\hline\hline
HD\,85628\,B&$147.57796550484$&$-66.11477795490$&$\,\,7.38_{-0.15}^{+0.23}$&$744$&$0.675_{-0.025}^{+0.016}$&$\,\,4282_{-104}^{+66}$ & \texttt{BPRP PRI}\\\hline
TOI\,717\,B&$147.98872815946$&$\,\,\,\,\,\,\,2.11708155887$&$\,\,9.72_{-0.13}^{+0.02}$&$2275$&$0.398_{-0.003}^{+0.017}$&$3519_{-3}^{+25}$ & \texttt{2MA BPRP PRI}\\\hline
G\,196-3\,B&$151.08506490419$&$\,\,\,\,50.38228242112$&$17.41_{-0.49}^{+0.48}$&$350$&$0.032_{-0.002}^{+0.002}$&$1987_{-93}^{+94}$ & \texttt{BD 2MA BPRP}\\\hline
LTT\,3780\,B&$154.64934761666$&$-11.71834621199$&$12.27_{-0.26}^{+0.34}$&$347$&$0.149_{-0.014}^{+0.014}$&$3127_{-58}^{+43}$ & \texttt{2MA BPR PPRI}\\\hline
MASCARA-3\,\,B&$161.90924448675$&$\,\,\,\,71.65515762432$&$\,\,7.83_{-0.28}^{+0.12}$&$200$&$0.625_{-0.013}^{+0.030}$&$\,\,4074_{-53}^{+126}$ & \\\hline
2M\,J1101-7732\,B&$165.33045751133$&$-77.54374799758$&$12.48_{-0.14}^{+0.14}$&$264$&$0.019_{-0.001}^{+0.001}$&$2574_{-21}^{+21}$ & \texttt{BD}\\\hline
WASP-175\,B&$166.31900970217$&$-34.12073886500$&$\,\,5.46_{-0.11}^{+0.10}$&$3971$&$0.919_{-0.016}^{+0.018}$&$5290_{-60}^{+65}$ & \texttt{2MA BPR PPRI}\\\hline
CHXR\,73\,C&$166.56385863699$&$-77.63060598303$&$\,\,8.04_{-1.03}^{+1.03}$&$8808$&$0.235_{-0.112}^{+0.148}$&$\,\,3280_{-260}^{+295}$ & \texttt{2MA BPRP}\\\hline
GJ\,414\,B&$167.76359381364$&$\,\,\,\,30.44392150823$&$\,\,8.06_{-0.57}^{+0.32}$&$406$&$0.587_{-0.034}^{+0.056}$&$\,\,3971_{-110}^{+249}$ & \texttt{2MABPRP PRI}\\\hline
HD\,97334\,BC$^2$&$168.10555672848$&$\,\,\,\,35.80289455668$&$18.16_{-0.19}^{+0.04}$&$2036$&$0.028_{-0.001}^{+0.001}$&$1845_{-8}^{+36}$ & \texttt{BD 2MA BPRP}\\\hline
HD\,233832\,B&$171.51750545565$&$\,\,\,\,50.37622350226$&$\,\,8.64_{-0.21}^{+0.11}$&$290$&$0.531_{-0.012}^{+0.022}$&$3775_{-36}^{+68}$ & \texttt{2MA BPRP PRI}\\\hline
2M\,J1155-7919\,B&$178.76276748600$&$-79.32082818366$&$14.48_{-0.06}^{+0.02}$&$582$&$0.016_{-0.001}^{+0.001}$&$2366_{-2}^{+10}$ & \texttt{BD 2MA BPRP}\\\hline
NGTS-5\,B&$221.06499706987$&$\,\,\,\,\,\,\,5.60206193982$&$\,\,9.67_{-0.11}^{+0.12}$&$8323$&$0.405_{-0.016}^{+0.014}$&$3530_{-20}^{+22}$ & \texttt{2MA BPRP}\\\hline
2M\,J1450-7841\,B&$222.67064914833$&$-78.69407266366$&$15.35_{-0.51}^{+0.51}$&$387$&$0.031_{-0.003}^{+0.005}$&$\,\,2328_{-100}^{+114}$ & \texttt{BD 2MA BPRP}\\\hline
WASP-189\,B&$225.68920534342$&$\,\,\,-3.03062676806$&$\,\,9.12_{-0.16}^{+0.23}$&$942$&$0.479_{-0.030}^{+0.021}$&$3646_{-48}^{+33}$ & \texttt{2MA BPRP PRI}\\\hline
HIP\,73990\,D&$226.82470360183$&$-29.49738818566$&$\,\,8.33_{-0.64}^{+0.45}$&$5234$&$0.470_{-0.098}^{+0.151}$&$\,\,3576_{-107}^{+191}$ & \texttt{2MA BPRP PRI}\\\hline
TOI\,905\,B&$227.66059753260$&$-71.36174373792$&$10.93_{-0.14}^{+0.17}$&$358$&$0.251_{-0.016}^{+0.014}$&$3340_{-25}^{+21}$ & \\\hline
2M\,1510\,B&$227.69777941730$&$-28.30671294175$&$15.13_{-0.14}^{+0.03}$&$249$&$0.033_{-0.001}^{+0.001}$&$2376_{-6}^{+31}$ & \texttt{BD 2MA BPRP}\\\hline
$\beta$\,Cir\,B&$229.33920903332$&$-58.85886078926$&$16.91_{-0.23}^{+0.16}$&$6187$&$0.063_{-0.002}^{+0.002}$&$2159_{-48}^{+47}$ & \texttt{BD 2MA BPRP}\\\hline
KELT-23\,B&$232.14912993468$&$\,\,\,\,66.35793842725$&$\,\,9.94_{-0.10}^{+0.05}$&$575$&$0.368_{-0.007}^{+0.014}$&$3487_{-7}^{+15}$ & \texttt{2MA BPRP PRI}\\\hline
K2-290\,C&$234.85788601749$&$-20.20202286624$&$10.51_{-0.26}^{+0.27}$&$3094$&$0.292_{-0.026}^{+0.033}$&$3404_{-41}^{+38}$ & \texttt{2MA BPRP***}\\\hline
GQ\,Lup\,C&$237.30537201266$&$-35.65336961027$&$\,\,9.70_{-0.34}^{+0.42}$&$2446$&$0.075_{-0.015}^{+0.016}$&$2897_{-33}^{+14}$ & \texttt{2MA BPRP}\\\hline
HIP\,77900\,B&$238.62692763313$&$-27.33270712891$&$13.48_{-0.16}^{+0.05}$&$3374$&$0.020_{-0.001}^{+0.001}$&$2519_{-8}^{+22}$ & \texttt{BD BPRP}\\\hline
USco\,1602-2401\,B$^3$&$240.71312906858$&$-24.03074240242$&$\,\,8.69_{-0.12}^{+0.12}$&$1038$&$0.238_{-0.017}^{+0.016}$&$3280_{-27}^{+25}$ & \texttt{2MA BPRP PRI}\\
\hline\hline
\end{tabular}
\end{table*}
\end{landscape}

\setcounter{table}{3}
\begin{landscape}
\begin{table*}
\caption{continued}
\begin{tabular}{lccccccc}
\hline\hline

Companion & $\alpha$   & $\delta$   & $M_{G}$ & $sep$ & $mass$        &  $T_{eff}$ & Flags \\
          & [$^\circ$] & [$^\circ$] & [mag]   & [au]  & [$M_{\odot}$] &  [K]       & \\
\hline\hline
HIP\,79098\,C&$242.20150369275$&$-23.68925441843$&$\,\,9.82_{-0.15}^{+0.24}$&$9555$&$0.159_{-0.020}^{+0.013}$&$3176_{-40}^{+24}$ & \texttt{2MA BPRP PRI ***}\\\hline
USco\,1610-1913\,B&$242.63466253957$&$-19.21910506083$&$12.71_{-0.39}^{+0.39}$&$777$&$0.025_{-0.003}^{+0.003}$&$2615_{-47}^{+47}$ & \texttt{BD 2MA BPRP}\\\hline
USco\,1612-1800\,B&$243.20402838620$&$-18.01380384763$&$12.50_{-0.39}^{+0.39}$&$504$&$0.026_{-0.003}^{+0.003}$&$2642_{-48}^{+48}$ & \texttt{BD BPRP}\\\hline
ROXs\,12\,C&$246.61560598363$&$-25.45695531769$&$\,\,7.28_{-1.00}^{+1.00}$&$5095$&$0.474_{-0.181}^{+0.259}$&$\,\,3657_{-289}^{+343}$ & \texttt{2MA BPR PPRI}\\\hline
HATS-48\,B&$288.66902407326$&$-59.57941400923$&$11.71_{-0.19}^{+0.37}$&$1444$&$0.181_{-0.022}^{+0.013}$&$3219_{-60}^{+30}$ & \texttt{2MA BPRP}\\\hline
GJ\,752\,\,B&$289.23745710752$&$5.14456304678$&$14.04_{-0.28}^{+0.17}$&$446$&$0.098_{-0.002}^{+0.005}$&$2785_{-36}^{+60}$ & \texttt{2MA BPRP PRI}\\\hline
HD\,181234\,B&$290.00108472870$&$-9.32417966740$&$10.46_{-0.20}^{+0.07}$&$247$&$0.297_{-0.007}^{+0.026}$&$3412_{-11}^{+29}$ & \texttt{2MA BPRP PRI}\\\hline
Wendelstein-1\,B&$299.04791069915$&$17.56797448132$&$11.43_{-0.47}^{+0.43}$&$3632$&$0.202_{-0.030}^{+0.047}$&$3264_{-70}^{+72}$ & \texttt{2MA BPRP}\\\hline
2M\,J2126-81\,B&$321.71158783493$&$-81.67526360458$&$17.84_{-0.43}^{+0.05}$&$7427$&$0.020_{-0.004}^{+0.001}$&$1851_{-77}^{+41}$ & \texttt{BD 2MA BPRP}\\\hline
TOI\,132\,B&$338.40325545731$&$-43.44166603901$&$12.21_{-0.16}^{+0.14}$&$3231$&$0.152_{-0.006}^{+0.009}$&$3137_{-24}^{+27}$ & \texttt{2MA BPRP}\\\hline
NGTS-7\,B&$352.52202473338$&$-38.97006605140$&$\,\,9.32_{-0.50}^{+0.50}$&$156$&$0.384_{-0.091}^{+0.097}$&$\,\,3495_{-97}^{+102}$ & \\\hline
DS\,Tuc\,B&$354.91457052896$&$-69.19458723114$&$\,\,5.86_{-0.24}^{+0.10}$&$237$&$0.834_{-0.027}^{+0.060}$&$\,\,5040_{-79}^{+180}$ & \texttt{2MA BPRP PRI}\\\hline
1RXS\,J2351+3127\,C&$357.93142859121$&$31.45083817280$&$\,\,9.61_{-0.43}^{+0.04}$&$5427$&$0.394_{-0.007}^{+0.057}$&$3522_{-8}^{+98}$ & \texttt{2MA BPRP PRI}\\
\hline\hline\\
\end{tabular}

\textbf{\underline{Table Footnotes:}}

$^1$: 2M\,0441+23\,B is a close brown dwarf companion of 2M\,0441+23\,A.

$^2$: HD\,97334\,BC is a binary brown dwarf system.

$^3$: The brown dwarf USco\,1602-2401\,B was detected by \cite{aller2013} and its possible companionship to USco\,1602-2401\,A, was revealed with photometry and follow-up spectroscopy, which was finally
proven in this study with the Gaia DR2 astrometry of the companion, i.e. confirmation of equidistance, and common proper motion, as well as test for gravitational stability.
USco\,1602-2401\,B is one of 14 reported substellar companions, detected by Gaia, which were also characterized in this study using their Gaia DR2 astro- and photometry. In general, the derived mass of these
substellar companions agrees well with the mass given in the literature, with a deviation of only a few $M_{Jup}$, on average. In contrast, for USco\,1602-2401\,B \cite{aller2013} derived a mass of $41^{+20}_{-13}\,M_{Jup}$ at an age of 5\,Myr ($47^{+20}_{-18}\,M_{Jup}$ at 10\,Myr) adopting a distance of about 145\,pc and no extinction. With the Gaia DR2 parallax and the Starhorse extinction estimate of the primary star and the G-band photometry of the companion we obtained a significantly higher mass of $0.238^{+0.016}_{-0.017}\,M_{\odot}$ at 5\,Myr ($0.309^{+0.020}_{-0.019}\,M_{\odot}$ at 10\,Myr). Adopting $A_G=0$\,mag yields a mass of the companion of $0.071^{+0.001}_{-0.001}\,M_{\odot}$ for 5\,Myr, and $0.104^{+0.001}_{-0.001}\,M_{\odot}$ for 10\,Myr, respectively. Therefore, we classify this companion here as low-mass star.

\end{table*}
\end{landscape}

\begin{table*} \caption{List of all detected companions, whose differential proper motion $\mu_{rel}$ exceeds their estimated escape velocity $\mu_{esc}$.}
\begin{center}
\begin{tabular}{lccc}
\hline\hline
Companion	     & $\mu_{rel}$    & $\mu_{esc}$      &\\
               	 & $[$mas/yr$]$   & $[$mas/yr$]$     &\\
\hline\hline
51\,Eri\,B\,(SB) & $19.04\pm0.71$ & $11.298\pm0.144$ & ***\\
HIP\,38594\,B    & $ \,\,\,6.44\pm0.24$ & $ \,\,\,4.965\pm0.082$ &\\
TOI\,905\,B	     & $ \,\,\,7.67\pm0.76$ & $ \,\,\,3.090\pm0.144$ &\\
\hline\hline
\end{tabular}
\end{center}
\label{Tab_gravtest}
\end{table*}

\section{Figures}

\end{document}